\documentclass[aps,prd,twocolumn,superscriptaddress,showpacs,preprintnumbers,nofootinbib,notitlepage]{revtex4-2}
\usepackage{graphicx}
\usepackage{bm}
\usepackage[utf8]{inputenc}
\usepackage[hyperfootnotes=false]{hyperref}
\usepackage{multirow}
\usepackage{amsmath}
\usepackage{amsfonts}
\usepackage{amssymb}
\usepackage{color}
\usepackage{xcolor}
\usepackage{xspace}
\usepackage{braket}
\usepackage{units}
\usepackage{slashed}
\usepackage[normalem]{ulem}
\usepackage{float}
\usepackage{tablefootnote}
\usepackage{threeparttable}

\graphicspath{ {plots/} }

\newcommand{\be}{\begin{equation}}
\newcommand{\ee}{\end{equation}}

\newcommand{\mevnospace}{\ensuremath{{\mathrm{\,Me\kern -0.1em V}}}}
\newcommand{\gevnospace}{\ensuremath{{\mathrm{\,Ge\kern -0.1em V}}}}
\newcommand{\tevnospace}{\ensuremath{{\mathrm{\,Te\kern -0.1em V}}}}

\usepackage{mfirstuc} 
\newcommand{\addReviewer}[2]{
  \expandafter\newcommand\csname #1\endcsname[1]{{\bf \color{#2} \capitalisewords{#1}:\,##1}}
  \expandafter\newcommand\csname #1cor\endcsname[2]{{\color{#2} \capitalisewords{#1}:\,\st{##1}{\,\bf ##2}}}
  \expandafter\newcommand\csname #1color\endcsname{\,#2}
}

\usepackage{tikz, marginnote} 
\newcommand{\checkedby}[1]{
\ifdefined\CROSSCHECKS
  \marginnote{
    \begin{tikzpicture}
      \foreach \x [count=\xi] in {#1} {
         \node[shape=circle,inner sep=0mm,
         minimum size=2mm,
         fill=\csname \x color\endcsname] at (\xi*3mm,0) {};
       }
    \end{tikzpicture}
  }
\else
\fi
}

\usepackage{soul,color}
\definecolor{chromeyellow}{rgb}{1.0, 0.65, 0.0}
\definecolor{DodgeBlue}{rgb}{0.118, 0.565,1.000}
\definecolor{asparagus}{rgb}{0.53, 0.66, 0.42}
\definecolor{cadmiumgreen}{rgb}{0.0, 0.42, 0.24}

\addReviewer{AR}{asparagus}
\addReviewer{JR}{red}
\addReviewer{JRE}{blue}

\newcommand{\ucm}{Departamento de F\'isica Te\'orica and IPARCOS, 
Universidad Complutense de Madrid, 
E-28040 Madrid, Spain}

\begin{document}

\preprint{IPARCOS-UCM-26-036}

\title{Poles in $\pi N$ scattering from forward dispersion relations and revised total cross-section data}
\author{J.~R.~Pel\'aez}
\email{jrpelaez@ucm.es}\affiliation{\ucm}
\author{P. Rab\'an}
\email{praban@ucm.es}\affiliation{\ucm}
\author{J.~Ruiz de Elvira}
\email{jacobore@ucm.es}\affiliation{\ucm}

\begin{abstract}
We present a model-independent calculation of $\pi N$ forward dispersion relations and their analytic continuation to the complex plane, using a revised set of $\pi^\pm p$ total cross-section data up to 3 GeV, and Regge asymptotics above. Up to that energy, we find four stable poles for each isospin combination. The lightest pole in the $I=3/2$ channel corresponds to the $\Delta(1232)$ resonance, while the lightest in the $I=1/2$ channel corresponds to the Roper resonance, even though the latter is imperceptible in the data.
We extract their pole parameters and the 
parameter difference between the $\Delta^0$ and $\Delta^{++}$, without relying on a partial-wave analysis. The remaining poles cannot be identified with a single resonance each. They are not artifacts but the combined effect of multiple resonances unresolved by total cross-section data alone. Finally, we write sum rules relating the residues of these constituent resonances to the residues of the poles extracted from the dispersive representation. 
\end{abstract}

\maketitle

\section{Introduction}

The spectrum of nucleon excitations encodes fundamental information on the non-perturbative dynamics of QCD. Resonances are identified with poles of the scattering amplitude in the complex-energy plane on unphysical Riemann sheets, whose positions and residues provide the most model-independent characterization of the state. In practice, however, the extraction of such parameters requires a controlled analytic continuation from the physical region, a task that is notoriously challenging for excited baryons.

Pion--nucleon scattering constitutes one of the primary channels for the study of baryon resonances. As the lightest meson-baryon reaction, the $\pi N$ system provides access to baryon excitations through a simple two-body process with well-measured total cross-sections~\cite{ParticleDataGroup:2026aaa}, and has long served as a
benchmark for resonance spectroscopy~\cite{Penner:2002ma,Arndt:2003if,Arndt:2006bf,Workman:2012hx,Anisovich:2011fc,Ronchen:2012eg,Kamano:2013iva}.
Standard studies extract baryon pole parameters from partial-wave analyses of differential cross-section data. While these methods have been highly successful, the extracted pole parameters usually depend on the choice of parametrization and the analytic continuation procedure and inherit systematic uncertainties associated with the partial-wave reconstruction, truncation, etc. Dispersive methods offer a complementary and largely model-independent approach, exploiting analyticity, unitarity, and crossing symmetry to constrain the amplitude without relying on a specific functional form.

The most rigorous dispersive framework for $\pi N$ scattering is
provided by Roy--Steiner equations~\cite{Ditsche:2012fv,Hoferichter:2012wf,Hoferichter:2015dsa,Hoferichter:2015tha,Hoferichter:2015hva,Hoferichter:2016ocj,Hoferichter:2016duk,Siemens:2016jwj,RuizdeElvira:2017stg,Hoferichter:2018zwu,Hoferichter:2023ptl,Hoferichter:2023mgy}, a
system of partial-wave dispersion relations that combines the
constraints from analyticity, unitarity, and crossing symmetry. This framework has recently been used to extract the pole parameters of the $\Delta(1232)$ and Roper resonances in a fully controlled manner~\cite{Hoferichter:2023mgy,RuizdeElvira:2025lwg}. The present work adopts a complementary strategy: rather than starting from partial-wave amplitudes, we reconstruct forward $\pi N$ scattering amplitudes directly from $\pi^\pm p$ total cross-sections via the optical theorem and forward dispersion relations (FDRs), and investigate how much information on the baryon resonance spectrum can be extracted from data on total cross-sections. Since such data do not reach the region very near threshold, only there will we make use of other input, but with little influence on the poles we find in the resonance region. 

A prerequisite for this analysis is the use of a reliable dataset of $\pi^\pm p$ total cross-sections. Comparing the compilation of the Review of Particle Physics (RPP)~\cite{ParticleDataGroup:2026aaa} with the original experimental publications and dedicated data reanalyses~\cite{Giacomelli:1970,Schopper:1988vrx}, here we identify several possible improvements upon the RPP compilation. These include: i) the systematic use of the charged-pion instead of the neutral-pion mass when converting different kinematic variables, following the original references; ii) the inclusion of additional data points or sets; iii) updated corrections of central values and uncertainties. 
We thus fit this revised dataset with a model-independent parametrization, which is then
corrected for electromagnetic and isospin-breaking effects, following Refs.~\cite{Arndt:2003if,Arndt:2006bf}.

The corrected fit just described provides our input for the dispersive reconstruction of the forward amplitudes up to $\sqrt{s}=3\,\mathrm{GeV}$. As input for the dispersive integrals above that energy, we use the Regge parametrization of high-energy data in the RPP~\cite{ParticleDataGroup:2016lqr}. 
In this work, for the first time, we analytically continue the output of the FDRs into the complex-energy plane by means of continued fractions~\cite{Schlessinger:1968}.
This approach, called the $\mathrm{FDR}_{C_N}$ method, has been previously applied to meson-meson scattering in~\cite{Pelaez:2022qby,Pelaez:2025jrn} to determine the existence and parameters of light-meson resonances. 
We find four stable poles in each of the $I=1/2$ and $I=3/2$ FDRs below 3 GeV.
As could be expected from its conspicuous peak in the total cross-section data, the lightest pole in the $I=3/2$ channel corresponds to the $\Delta(1232)$ resonance. The lightest pole found in the $I=1/2$ amplitude corresponds to the Roper or $N(1440)$ resonance, although its presence is imperceptible in the total cross-section data. Thus, our results constitute the first model-independent determination of its existence and parameters without relying on a partial-wave analysis.

The remaining six poles do not correspond to individual resonances, raising the question of whether they are merely artifacts of the approach. We show that they are not but instead arise from the combined contributions of the established $N$- and $\Delta$-resonances in that region, which are not resolved individually by the total cross-section data. We establish this connection from sum rules that relate the residue of each dispersive pole to those of the underlying resonances that generate it. Finally, we provide a model-independent value for the isospin-breaking splitting between the $\Delta^{0}$ and $\Delta^{++}$ charged states, once again avoiding the use of partial-wave analyses.

The paper is organized as follows. In Sect.~\ref{sec:amp}, we review the notation for $\pi N$ scattering amplitudes, the forward dispersion relations, and the isospin and electromagnetic corrections applied to the data. Section~\ref{sec:method} describes the revision of the experimental dataset, the fit to the $\pi^\pm p$ total cross-sections, and the resulting dispersive amplitudes. The extraction of resonance poles with the $\mathrm{FDR}_{C_N}$ method and the corresponding results are presented in Sect.~\ref{sec:results}, including a study of isospin-breaking effects in the $\Delta(1232)$ sector. Section~\ref{sec:discussion} contains the summary and conclusions.

\section{$\pi N$ scattering amplitudes and forward dispersion
relations}\label{sec:amp}

In this section, we briefly review the $\pi N$ scattering amplitudes and fix the notation, following the conventions of H\"ohler~\cite{Hohler:1984ux}. We then recall the standard forward dispersion relations for the $D^\pm$ amplitudes~\cite{Martin:1970xx,Itzykson:1980rh,Hohler:1984ux,Gasser:1988jt} and discuss the electromagnetic and isospin-breaking corrections needed to obtain the purely hadronic and isospin-symmetric input for the dispersive integrals.

\subsection{$\pi N$ Kinematics and notation}

We consider the reaction
\begin{equation}
\pi^a(q)+N(p)\to \pi^b(q')+N(p'),
\end{equation}
where $a,b$ denote pion isospin indices and $N$ the nucleon SU(2) doublet. The
$\pi N$ scattering amplitude can be decomposed into isospin-even and
isospin-odd components as
\begin{align}
T^{ba}_{fi}(s,t)
&=\frac{1}{2}\{\tau^b,\tau^a\}\,T^+_{fi}(s,t)
+\frac{1}{2}[\tau^b,\tau^a]\,T^-_{fi}(s,t) \nonumber\\
&=\delta^{ba}T^+_{fi}(s,t)+i\epsilon^{bac}\tau^c\,T^-_{fi}(s,t),
\end{align}
where $\tau^a$ are the Pauli matrices and $s=(p+q)^2$, $t=(q-q')^2$, and
$u=(p-q')^2$ are the usual Mandelstam variables. The most general
Lorentz-invariant and parity-conserving decomposition of the amplitudes
$T^\pm_{fi}$ reads
\begin{align}
T^\pm_{fi}(s,t)
&=\bar u_f(p')\left\{
A^\pm(\nu,t)+\frac{\slashed q'+\slashed q}{2}\,B^\pm(\nu,t)
\right\}u_i(p) \nonumber\\
&=\bar u_f(p')\left\{
D^\pm(\nu,t)-\frac{[\slashed q',\slashed q]}{4m_N}\,B^\pm(\nu,t)
\right\}u_i(p),
\end{align}
where $m_N$ and $M_\pi$ denote the nucleon and pion masses, respectively, and
the spinors are normalized as $\bar u u=2m_N$. The amplitude $D^\pm$ is defined by
\begin{equation}
D^\pm(\nu,t)=A^\pm(\nu,t)+\nu\,B^\pm(\nu,t),
\qquad
\nu=\frac{s-u}{4m_N},
\end{equation}
with
\begin{equation}
u=2m_N^2+2M_\pi^2-s-t.
\end{equation}

The invariant amplitudes satisfy the crossing relations
\begin{align}
A^\pm(\nu,t)&=\pm A^\pm(-\nu,t), &
B^\pm(\nu,t)&=\mp B^\pm(-\nu,t), \nonumber\\
D^\pm(\nu,t)&=\pm D^\pm(-\nu,t).
\label{eq:D_crossing}
\end{align}

It is convenient to project onto states of definite isospin. For any of the invariant amplitudes $\mathcal A\in\{A,B,D\}$, one has
\begin{equation}
\bra{I',I_3'}\mathcal A\ket{I,I_3}
=
\mathcal A(I)\,\delta_{II'}\delta_{I_3I_3'}
\equiv
\mathcal A^I\,\delta_{II'}\delta_{I_3I_3'}.
\end{equation}
This yields the relations
\begin{align}
\mathcal A_+
&\equiv \mathcal A(\pi^+ p\to \pi^+ p)
=\mathcal A^+-\mathcal A^-
=\mathcal A^{3/2}, \nonumber\\
\mathcal A_-
&\equiv \mathcal A(\pi^- p\to \pi^- p)
=\mathcal A^++\mathcal A^-
=\frac{1}{3}\left(2\mathcal A^{1/2}+\mathcal A^{3/2}\right),
\label{eq:amplitude_relations}
\end{align}
which relate the isospin amplitudes to the experimentally measured $\pi^\pm p$ channels.

Finally, for the forward physical amplitudes
$D_\pm(s,0)$, the optical theorem implies
\begin{equation}
\operatorname{Im} D_\pm(s,0)=k_{\mathrm{lab}}\,\sigma_\pm,
\label{eq:optical_theorem}
\end{equation}
where $\sigma_\pm\equiv \sigma(\pi^\pm p)$ are the total cross-sections and
\begin{equation}
\omega_{\mathrm{lab}}=\frac{s-m_N^2-M_\pi^2}{2m_N},
\qquad
k_{\mathrm{lab}}=\sqrt{\omega_{\mathrm{lab}}^2-M_\pi^2},
\end{equation}
denote the pion energy and momentum in the laboratory frame, respectively.
Equation~\eqref{eq:optical_theorem} provides the imaginary parts of $D^\pm$ entering the forward dispersion relations discussed in the following subsection.

\subsection{$\pi N$ forward dispersion relations for the $D^\pm$ amplitudes}

To simplify the nucleon-pole terms and to compare directly with the literature, we first introduce the so-called pseudovector-Born-term-subtracted amplitudes
\begin{align}
\bar D^+(s,t)=&\,D^+(s,t)-\frac{g^2}{m_N}
-\nu\,g^2\left(\frac{1}{m_N^2-s}-\frac{1}{m_N^2-u}\right) \nonumber \\
=&\,D^+(s,t)-\frac{g^2}{m_N}\frac{\nu_B^2}{\nu_B^2-\nu^2}, \nonumber\\
\bar D^-(s,t)=&\,D^-(s,t)+\frac{\nu\,g^2}{2m_N^2}
-\nu\,g^2\left(\frac{1}{m_N^2-s}+\frac{1}{m_N^2-u}\right) \nonumber \\
=&\,D^-(s,t)+\frac{\nu\,g^2}{2m_N^2}\left(1-\frac{2m_N\nu_B}{\nu_B^2-\nu^2}\right),
\label{eq:D_bar}
\end{align}
where $\nu_B=\frac{2m_N^2-s-u}{4m_N}$, and $g$ is the $\pi N$ pseudoscalar coupling constant. We use the value
$g^2/4\pi=13.7\pm0.2$ obtained from the Goldberger--Miyazawa--Oehme sum rule~\cite{Goldberger:1955zza}, as quoted in~\cite{Baru:2010xn,Baru:2011bw}, used in the Roy--Steiner analysis~\cite{Hoferichter:2015hva} and recently updated in~\cite{Hoferichter:2023ptl}. This value is fully consistent with the more precise determination in~\cite{Reinert:2020mcu}.

The forward amplitudes $D^\pm$ satisfy dispersion relations at $t=0$ that follow from analyticity, unitarity, crossing symmetry, and the high-energy behavior of the total cross-sections.
Since $D^+$ is crossing even, a once-subtracted forward dispersion relation is sufficient~\cite{Hohler:1984ux} to ensure the convergence of the dispersive integrals. We choose the subtraction point at $\nu=0$, where the subtraction constant is fixed by the isospin-even S-wave scattering length $a_{0+}^+$. Writing the relation in terms of $k_{\rm lab}$, one obtains~\cite{Hohler:1984ux,Gasser:1988jt}
\begin{align}
\operatorname{Re}\bar D^+(k_{\rm lab})
&=4\pi\left(1+\frac{M_\pi}{m_N}\right)\bar a_{0+}^+\nonumber\\
&+\frac{k_{\rm lab}^2}{\pi}\,\mathcal P\!\!\int_0^\infty
\frac{\sigma_+(k'_{\rm lab})+\sigma_-(k'_{\rm lab})}
{k_{\rm lab}'^{\,2}-k_{\rm lab}^2}\,dk'_{\rm lab},
\label{eq:FDR_+}
\end{align}
where $\mathcal{P}$ denotes the principal value. Note that we have used~\eqref{eq:optical_theorem} to express the imaginary parts in terms of total cross-sections, and the subtraction constant is related to the scattering length as defined in~\cite{Gasser:1988jt}
\begin{equation}
4\pi(1+x)\bar a_{0+}^+
=
4\pi(1+x)a_{0+}^+
+\frac{g^2x^3}{M_\pi(4-x^2)},
\end{equation}
with $x=M_\pi/m_N$. 

The situation for the crossing-odd amplitude $D^-$ is different. Since $D^-(\nu,t)$ is odd under $s\leftrightarrow u$ crossing, one might expect the need for two subtractions. However, the Pomeranchuk theorem~\cite{Pomeranchuk:1958ged} ensures that
$\sigma_+(z)-\sigma_-(z)\to 0$ sufficiently fast as $z\to\infty$, so that a once-subtracted relation also converges. It is convenient to subtract at the
subthreshold point $\nu=0$, where $D^-(0,0)=0$, so writing the dispersion relation in
terms of the pion laboratory momentum $k_{\rm lab}$ one gets~\cite{Hohler:1984ux,Gasser:1988jt},
\begin{align}
\operatorname{Re}\bar D^-(k_{\rm lab})
&=
\frac{g^2\,\omega_{\rm lab}}{2m_N^2}\nonumber\\
&+\frac{\omega_{\rm lab}}{\pi}\,\mathcal{P}\!\!\int_0^\infty
\frac{k_{\rm lab}'^{\,2}}{\omega_{\rm lab}'}
\frac{\sigma_-(k_{\rm lab}')-\sigma_+(k_{\rm lab}')}{k_{\rm lab}'^{\,2}-k_{\rm lab}^2}\,dk_{\rm lab}',
\label{eq:FDR_-}
\end{align}
where we have used again~\eqref{eq:optical_theorem}.

Forward dispersion relations for these amplitudes are particularly useful for three reasons. First, in generic dispersion relations, the integrands contain the imaginary part of the amplitude, but in this case, the optical theorem~\eqref{eq:optical_theorem}
has allowed us to replace them with the $\pi^\pm p$ total cross-sections,
which are measurable quantities.
Second, $D^\pm$ have definite crossing properties under $s\leftrightarrow u$, which relate the left-hand cut to the physical right-hand cut and ensure that the resulting dispersion relations involve experimentally accessible total cross-sections.
Third, the crossing-even amplitude $D^+$ satisfies a strong positivity bound~\cite{Hohler:1984ux},
\begin{equation}
\left(\frac{\partial}{\partial t}\right)^n \operatorname{Im}D^+(s,0)\geq \left|\left(\frac{\partial}{\partial t}\right)^n \operatorname{Im}D^+(s,t) \right|,
\end{equation}
for $0\geq t\geq -4q_{\rm cm}^2$ and any non-negative integer $n$, where
\begin{equation}
q_{\rm cm}=
\frac{\lambda^{1/2}\!\left(s,m_N^2,M_\pi^2\right)}{2\sqrt{s}}
\end{equation}
is the center-of-mass momentum, with $\lambda(a,b,c)=a^2+b^2+c^2-2(a\,b+b\,c+c\,a)$. This property avoids cancellations in the dispersive integrals, and hence, it improves the numerical stability of the $D^+$ forward dispersion relation. No analogous positivity condition exists for $D^-$, which partly explains the larger uncertainties typically obtained for the crossing-odd amplitude.

Note that both $a_{0+}^+$ and the isospin-odd scattering length $a_{0+}^-$ are two of the few pieces of data that do not come from total cross-section experiments. They are needed not only due to the subtraction term, but also to describe the total cross-sections at threshold, as will be discussed in Sect.~\ref{sec:method}.  They can be determined precisely from pionic-atom data~\cite{Baru:2010xn,Baru:2011bw,Hoferichter:2016ocj,Hoferichter:2023ptl}, provided the
corresponding isospin-breaking corrections are taken into
account \cite{Gasser:2002am,Hoferichter:2009ez,Hoferichter:2009gn,Hoferichter:2012bz}. Throughout this work, we use the values quoted in~\cite{Hoferichter:2015hva}.
Changing their values well beyond their estimated uncertainties has a negligible effect on the poles we find in the FDRs.

\subsection{Electromagnetic and isospin-breaking corrections}
\label{sec:corrections}

The dispersive integrals in equations~\eqref{eq:FDR_+} and~\eqref{eq:FDR_-} require purely hadronic and isospin-symmetric total cross-sections as input, i.e., removing electromagnetic interactions and isospin-breaking effects. While most experimental datasets already subtract the pure Coulomb contribution and the factorized Coulomb--nuclear interference term, residual electromagnetic effects remain where the hadronic and Coulomb amplitudes cannot be cleanly separated. A standard framework for these corrections is the Nordita scheme~\cite{Tromborg:1976bi}, which provides electromagnetic corrections for the S- and P-wave contributions. The original Nordita corrections, however, are reliable only up to $\sqrt{s}\simeq1.47\,\mathrm{GeV}$, and one of the isospin-mixing terms was later found to be incorrect~\cite{Arndt:2003if}. We therefore adopt the electromagnetic corrections from~\cite{Arndt:2003if}, obtained from the SAID/GWU database~\cite{SAID} as the difference between the total cross-sections provided there and those reconstructed from the purely hadronic and isospin-symmetric partial waves. Below $\sqrt{s}\simeq1.47\,\mathrm{GeV}$, these corrections are consistent with the original Nordita prescription; above that energy, they incorporate additional contributions from higher partial waves.

Isospin-breaking effects require an analogous treatment, since the
dispersion relations are formulated in the isospin-symmetric limit.
The dominant correction arises from the mass and width splitting
between the $\Delta^0$ and $\Delta^{++}$ resonances, which we
incorporate following~\cite{Arndt:2006bf}. These corrections are
applied to the data used in our dispersive analysis. A separate
determination of the $\Delta^0-\Delta^{++}$ splitting, based on
the uncorrected physical total cross-sections, is presented in
Sect.~\ref{sec:IB_corr}.

\section{Dispersive analysis of $\pi N$ scattering from total cross-section data}\label{sec:method}

The forward dispersion relations derived in the previous section require hadronic and isospin-symmetric $\pi^\pm p$ total cross-sections as input. We therefore begin with a reassessment of the experimental database compiled in the RPP~\cite{ParticleDataGroup:2026aaa}, from which we construct a consistent set of total cross-section data. 
After applying the corrections discussed in Sect.~\ref{sec:corrections}, we determine parametrizations of the resulting hadronic and isospin-symmetric total cross-sections and evaluate the forward dispersion relations~\eqref{eq:FDR_+} and~\eqref{eq:FDR_-}.

\subsection{Revision of total cross-section data}\label{sec:data}

\begin{figure*}
\centering
\includegraphics[width=0.99\textwidth]{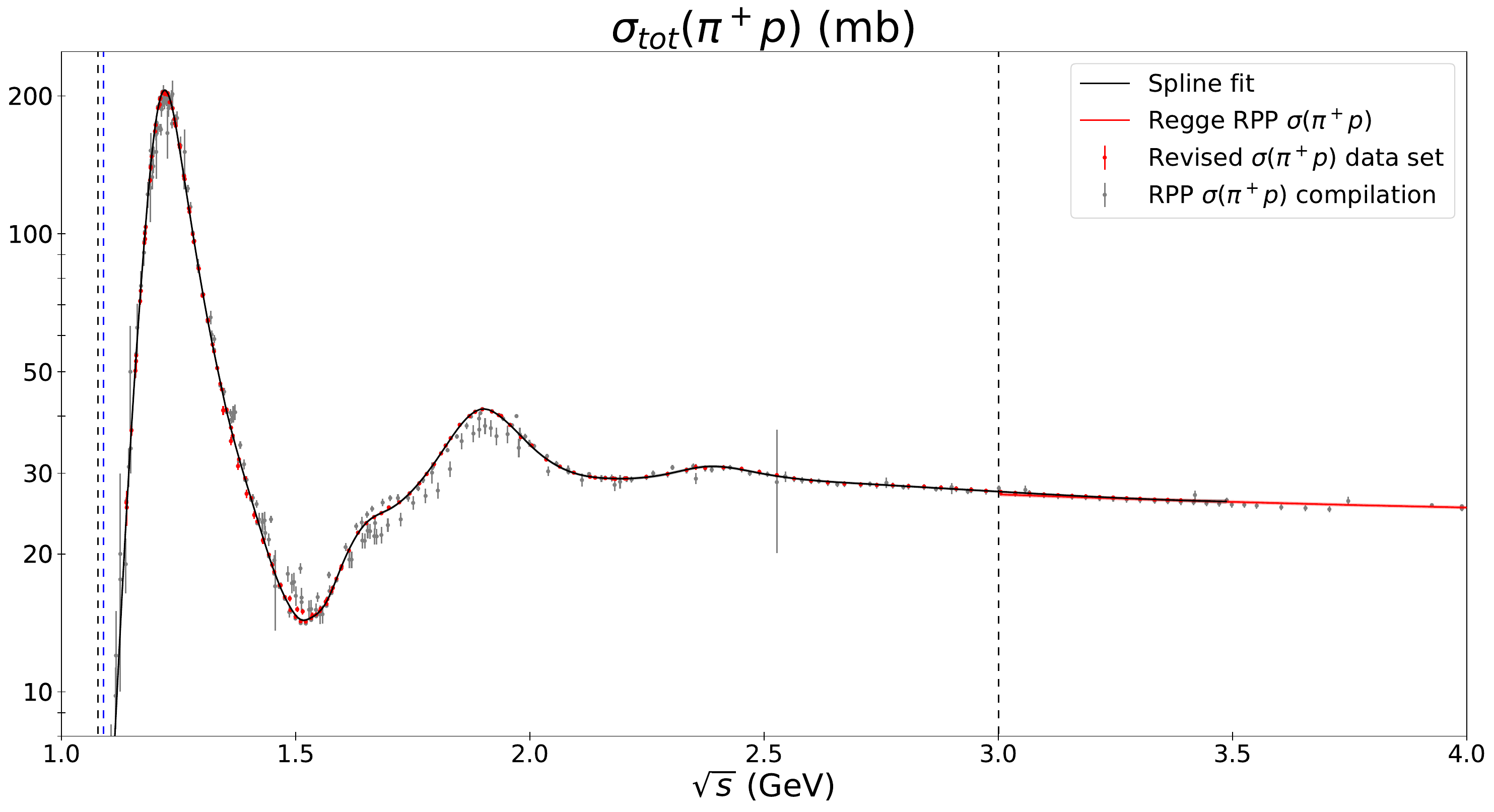}
\includegraphics[width=0.99\textwidth]{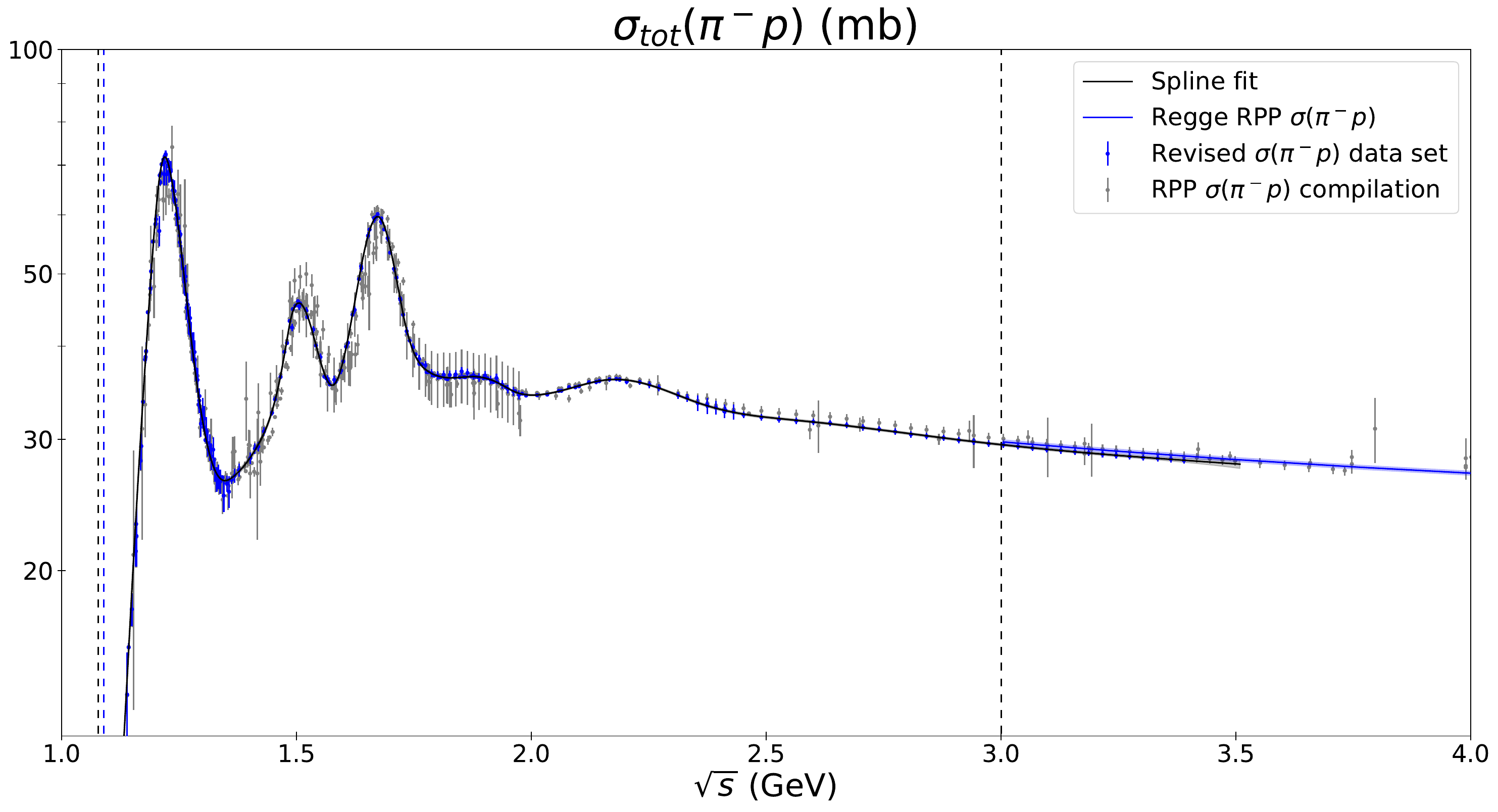}
\caption{Final parametrization of the $\pi^\pm p$ total cross-section data in the whole energy range. Colored points represent the revised dataset with the final normalizations, while gray points show the RPP~\cite{ParticleDataGroup:2026aaa} compilation. The first vertical dashed line stands for the $\pi N$ threshold, whereas the second one marks the maximum energy where the Roy--Steiner representation of~\cite{Hoferichter:2015hva} is used as input. Above the last vertical dashed line, the Regge representation of the RPP~\cite{ParticleDataGroup:2016lqr} is used as input for the integrals.}
\label{fig:cross-sections-entire}
\end{figure*}

Our analysis is based on the $\pi^\pm p$ total cross-section compilation provided in the RPP~\cite{ParticleDataGroup:2026aaa}, which contains a large number of data points from different datasets covering different energy ranges and with varying levels of precision, as seen in Fig.~\ref{fig:cross-sections-entire}. 
In this work, we have nevertheless revised the RPP compilation, tracing each dataset back to the original publications and to the dedicated reanalyses of Refs.~\cite{Giacomelli:1970,Schopper:1988vrx}. As a result, we have included missing measurements, have used the charged-pion mass systematically for kinematic conversions, and have revised and updated some central points as well as the assignment of systematic uncertainties. 

Since experiments with disproportionally large uncertainties have little impact on the total cross-section fit, we have removed them from our dataset, keeping just~\cite{Leonard:1954,Zinov:1960iwr,Diddens:1963zz, Devlin:1965zz,Bizard:1966yb,Citron:1966zz,Carter:1968zza, Bizard:1970ca,Carter:1971tj,Davidson:1972ky,Pedroni:1978it}. 

We have also excluded several of the oldest experiments~\cite{Leonard:1954,Diddens:1963zz,Devlin:1965zz, Bizard:1966yb} because they lack a detailed description of their treatment of electromagnetic effects discussed in Sect.~\ref{sec:corrections}. Otherwise, we could not implement such electromagnetic corrections consistently to all datasets.  Fortunately, their omission has a negligible impact on the final results, since more recent measurements provide both higher precision and broader coverage of the resonance region. Although not used as input to our fit, for completeness, let us note several issues in their compilation by the RPP. The $\pi^-p$ measurements of~\cite{Diddens:1963zz} are absent, whereas the corresponding $\pi^+p$ data are included. Originally, the Ref.~\cite{Devlin:1965zz} data were given in terms of the charged-pion kinetic energy, but were converted to laboratory momentum using the $\pi^0$ mass for the $\pi^-p$ data. In addition, the values of the systematic $\pi^+p$ uncertainties, originally in millibarns, are 
artificially reduced when interpreted as percentages. Concerning the \cite{Bizard:1966yb} dataset, several measurements at the same pion kinetic energy correspond to different running periods, but the RPP selected values do not always correspond to the most recent or precise determinations. Moreover, several $\pi^-p$ measurements at $T_\pi=330$, $375$, $425$, and $590~\mathrm{MeV}$ are absent from the RPP compilation.

Let us turn back to the datasets that we have included in our fit, namely those of Refs.~\cite{Zinov:1960iwr,Citron:1966zz,Carter:1968zza, Bizard:1970ca,Carter:1971tj,Davidson:1972ky,Pedroni:1978it}, which are shown in Fig.~\ref{fig:cross-sections-entire} as colored points.  Since they were reanalyzed in Refs.~\cite{Giacomelli:1970,Schopper:1988vrx}, whenever the central values or the uncertainties differ between the original works and the reanalyses, we have kept the latter.
Compared with the RPP compilation, our revision is particularly relevant for the central values of~\cite{Citron:1966zz}.  In the case of~\cite{Zinov:1960iwr}, it is the systematic error that changed in the reanalyses. In addition, we have used the charged-pion mass for the conversion from pion kinetic energy to laboratory momentum, instead of the neutral-pion mass used by the RPP. Moreover, the RPP adds a large systematic error to the original data of \cite{Bizard:1970ca}, which makes their total uncertainty artificially large, as seen in the 1.75--2 GeV region in the lower panel of Fig.~\ref{fig:cross-sections-entire}. However, the reanalysis in \cite{Schopper:1988vrx} provided a smaller total uncertainty, which we have adopted.

\subsection{Fit to $\pi^\pm p$ total cross-section data}

The $D^\pm$ amplitudes are parametrized up to $k_{\mathrm{lab}}=4.32\,\mathrm{GeV}$ ($\sqrt{s}=3\,\mathrm{GeV}$) using the revised $\pi^\pm p$ total cross-section data described in
the previous subsection. Very close to threshold, total cross-section data are absent. We could have just extrapolated our parametrization from the lowest data point to obtain the scattering lengths at threshold. However,  
we have also used input from the Roy--Steiner representation of~\cite{Hoferichter:2015hva} from threshold up to $k_{\mathrm{lab}}=64\,\mathrm{MeV}$. This tiny region is shown between dashed vertical lines in Fig.~\ref{fig:cross-sections-entire}.
This is the other instance of our use of input beyond total cross-section data. Using it or not has negligible effects on our results, and we have done it for almost cosmetic reasons to have an accurate fit in the very near-threshold region.

To avoid model dependencies, an order-three spline parametrization is constructed in several steps. First, a fit to data is used to identify a suitable set of the so-called spline knots, i.e., the matching points between splines. This is done using a smoothness parameter, ensuring that the main data structures are reproduced without fitting statistical fluctuations. On a second step, we introduce, as additional fit parameters, normalization factors for each dataset.
Then we follow the approach of \cite{RuizdeElvira:2017stg} for $\pi N$ scattering,
using the iterative fitting procedure of~\cite{Ball:2009qv} to avoid the D'Agostini bias~\cite{DAgostini:1993arp}.

Most of the resulting normalization factors and $\hat{\chi}^2\equiv \chi^2/n_{\rm data}$ for each dataset are of order one. An outstanding exception is the $\pi^-p$ data of Ref.~\cite{Davidson:1972ky}, which yield $\hat{\chi}^2\simeq79$.
Thus, following the consistency criteria of~\cite{Perez:2015pea}, we remove the $\pi^-p$ data of~\cite{Davidson:1972ky}, because it is dramatically inconsistent with the other six, as it has also been done in ~\cite{Pavan:1999cr,Pavan:private,Ericson:2000md}. After this exclusion, we repeat the fitting procedure and obtain our final result, whose normalizations and $\hat{\chi}^2$  
are collected in Table~\ref{tab:Norms}. All normalization factors remain even closer to unity for all experiments, indicating the overall consistency of the revised dataset. The quality of our fit to the Carter data set~\cite{Carter:1971tj} is significantly worse than for the other data sets, due to its very small experimental
uncertainties compared to the rest. Indeed, this reflects the known tensions among low-energy measurements in this channel~\cite{Pavan:1999cr,Pavan:private,Ericson:2000md}. Nevertheless, it provides the most precise coverage of the $\Delta(1232)$ peak region, where it constitutes an essential constraint on the parametrization.
 

\begin{table}[h]
\footnotesize
\renewcommand{\arraystretch}{1.6}
\setlength{\tabcolsep}{2mm}
\begin{threeparttable}
\begin{tabular}{ccccc}
\hline\hline
\multicolumn{5}{c}{Normalization factors $N$ and 
$\hat{\chi}^2$ values} \\
\hline
\multirow{2}{*}{Reference}
  & \multicolumn{2}{c}{$\sigma(\pi^+p)$}
  & \multicolumn{2}{c}{$\sigma(\pi^-p)$} \\
  & $N$ & $\hat{\chi}^2$ & $N$ & $\hat{\chi}^2$ \\
\hline
Zinov et al. 60~\cite{Zinov:1960iwr}
  & & & $1.013(11)$ & $0.9$ \\
Citron et al. 66~\cite{Citron:1966zz}
  & $1.002(2)$ & $0.07$ & $0.980(5)$ & $0.01$ \\
Carter et al.\ 68~\cite{Carter:1968zza}
  & $0.998(2)$ & $1.6$ & $0.993(2)$ & $0.7$ \\
Bizard et al.\ 70~\cite{Bizard:1970ca}
  & & & $1.012(3)$ & $1.05$ \\
Carter et al.\ 71~\cite{Carter:1971tj}
  & $0.992(2)$ & $3.0$ & $0.998(11)$ & $7.7$ \\
Davidson et al. 72~\cite{Davidson:1972ky}
  & $1.009(2)$ & $1.5$ & Excluded & $79$\,\tnote{(a)} \\
Pedroni et al. 78~\cite{Pedroni:1978it}
  & $0.998(1)$ & $1.6$ & $1.003(3)$ & $1.75$ \\
\hline\hline
\end{tabular}
\begin{tablenotes}
\footnotesize
\item[(a)] Before it is excluded from the fit.
\end{tablenotes}
\end{threeparttable}
\caption{Normalization factors and
$\hat{\chi}^2$ values for each dataset, obtained from the fit to the revised
$\pi^\pm p$ total cross-section data.
}
\label{tab:Norms}
\end{table}

\begin{figure*}[p]
\centering
\includegraphics[width=0.75\textwidth]{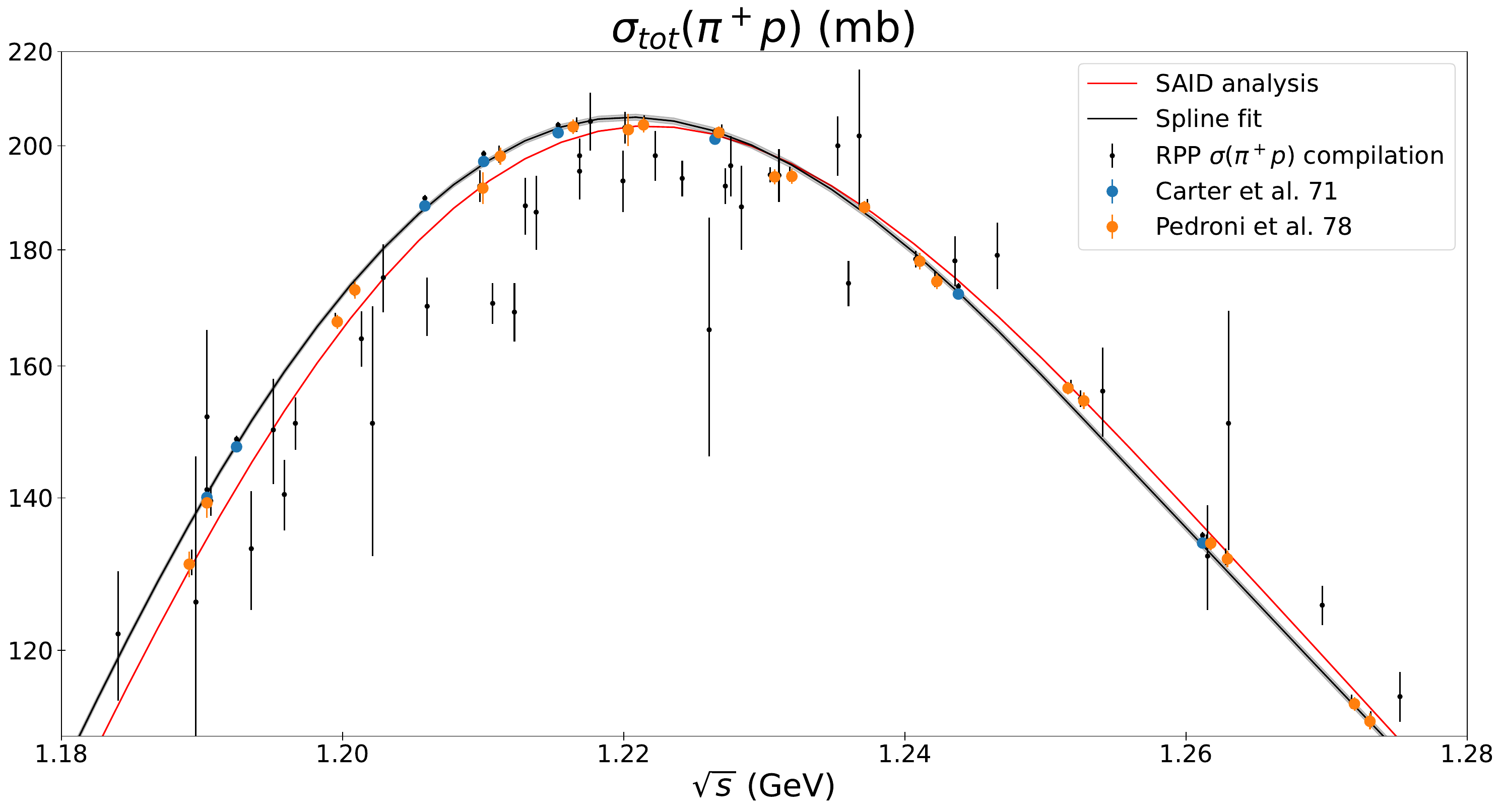}
\includegraphics[width=0.75\textwidth]{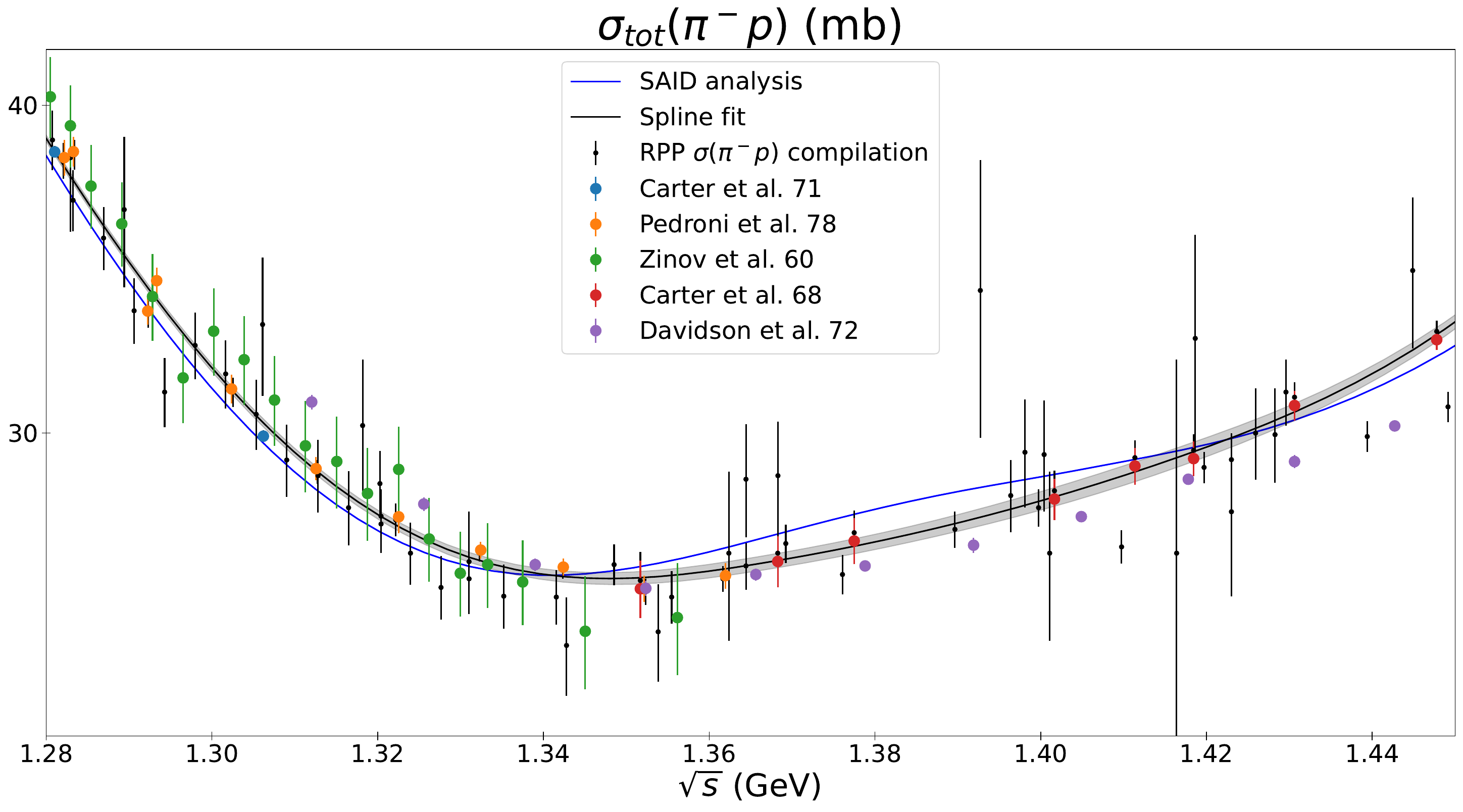}
\includegraphics[width=0.75\textwidth]{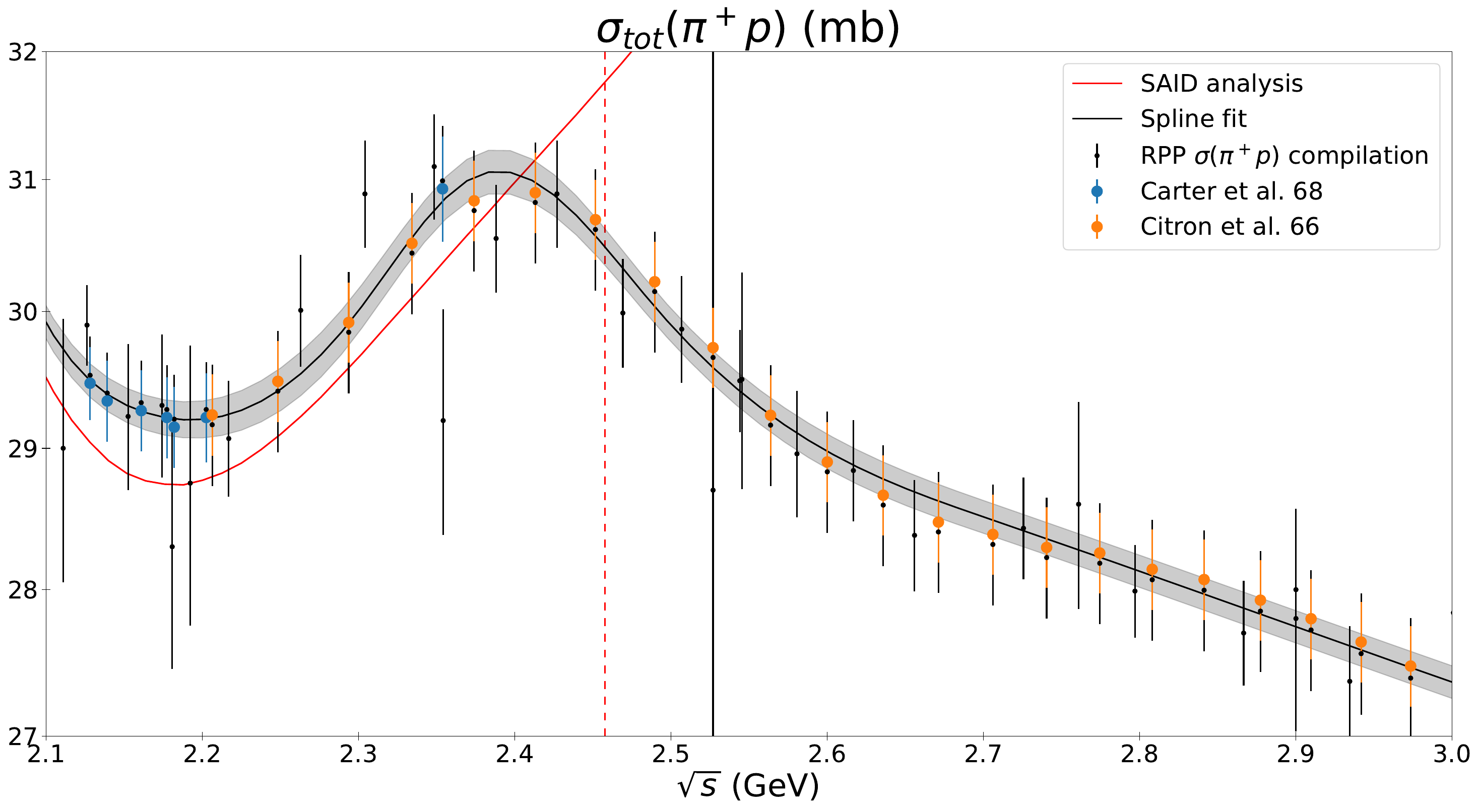}
\caption{Final parametrization of the $\pi^\pm p$ total cross-section data in three representative energy regions. Black points show the RPP~\cite{ParticleDataGroup:2026aaa} compilation; colored markers indicate the revised datasets after applying the fitted normalization factors. The SAID/GWU analysis~\cite{Workman:2012hx,SAID} is shown for comparison. The red vertical dashed line stands at the estimated validity range of the SAID/GWU parametrization \cite{Arndt:2006bf}.}
\label{fig:cross-sections}
\end{figure*}

The comparison between the final parametrization, the RPP compilation~\cite{ParticleDataGroup:2026aaa}, and the SAID/GWU partial-wave analysis~\cite{Workman:2012hx,SAID} is shown in Fig.~\ref{fig:cross-sections} for three representative energy regions. The SAID/GWU solution is of particular interest, as its input is based primarily on differential cross-sections and polarization observables, whereas ours relies primarily on $\pi^\pm p$ total cross-sections. In particular, among the high-precision total cross-section measurements considered here, only a limited subset is included in the SAID/GWU database~\cite{Zinov:1960iwr,Carter:1968zza,Bizard:1966yb}.\footnote{We thank I.~I.~Strakovsky and R.~L.~Workman for kindly providing their database.}

In the upper panel of Fig.~\ref{fig:cross-sections}, which shows the $\pi^+ p$ total cross-section in the $\Delta(1232)$ region, both analyses provide a qualitatively similar description of the data, although their central values are not fully compatible within uncertainties. This is also the energy region where the available total cross-section measurements provide the weakest constraints on the SAID/GWU solution. The middle panel shows $\sigma(\pi^- p)$ in the Roper region, where the agreement is generally better, except near $\sqrt{s}\simeq1.38\,\mathrm{GeV}$, where the SAID/GWU solution exhibits a recognizable enhancement associated with the $N(1440)$; no corresponding feature is apparent in the total cross-sections or in the present parametrization. The lower panel shows the highest-energy interval of the $\pi^+ p$ total cross-section. Both descriptions remain similar up to $\sqrt{s}\simeq2.4\,\mathrm{GeV}$, above which the SAID/GWU solution departs from both the experimental data and the present reconstruction. This is consistent with the SAID/GWU estimation~\cite{Arndt:2006bf} of $\sqrt{s}\approx 2.46$ GeV as their upper validity limit, which we have represented by a vertical red dashed line in the lower panel of Fig.~\ref{fig:cross-sections}.

\subsection{Dispersive amplitudes}

The dispersive integrals require input over the full energy range.
Up to $k_{\mathrm{lab}}=4.32\,\mathrm{GeV}$ ($\sqrt{s}=3\,\mathrm{GeV}$), we use the hadronic and isospin-symmetric total cross-sections from the parametrization described above, after applying the electromagnetic and isospin-breaking corrections of Sect.~\ref{sec:corrections}. For higher energies, we employ the Regge representation of the RPP~\cite{ParticleDataGroup:2016lqr}.

\begin{figure*}[ht]
\centering
\includegraphics[width=0.8\textwidth]{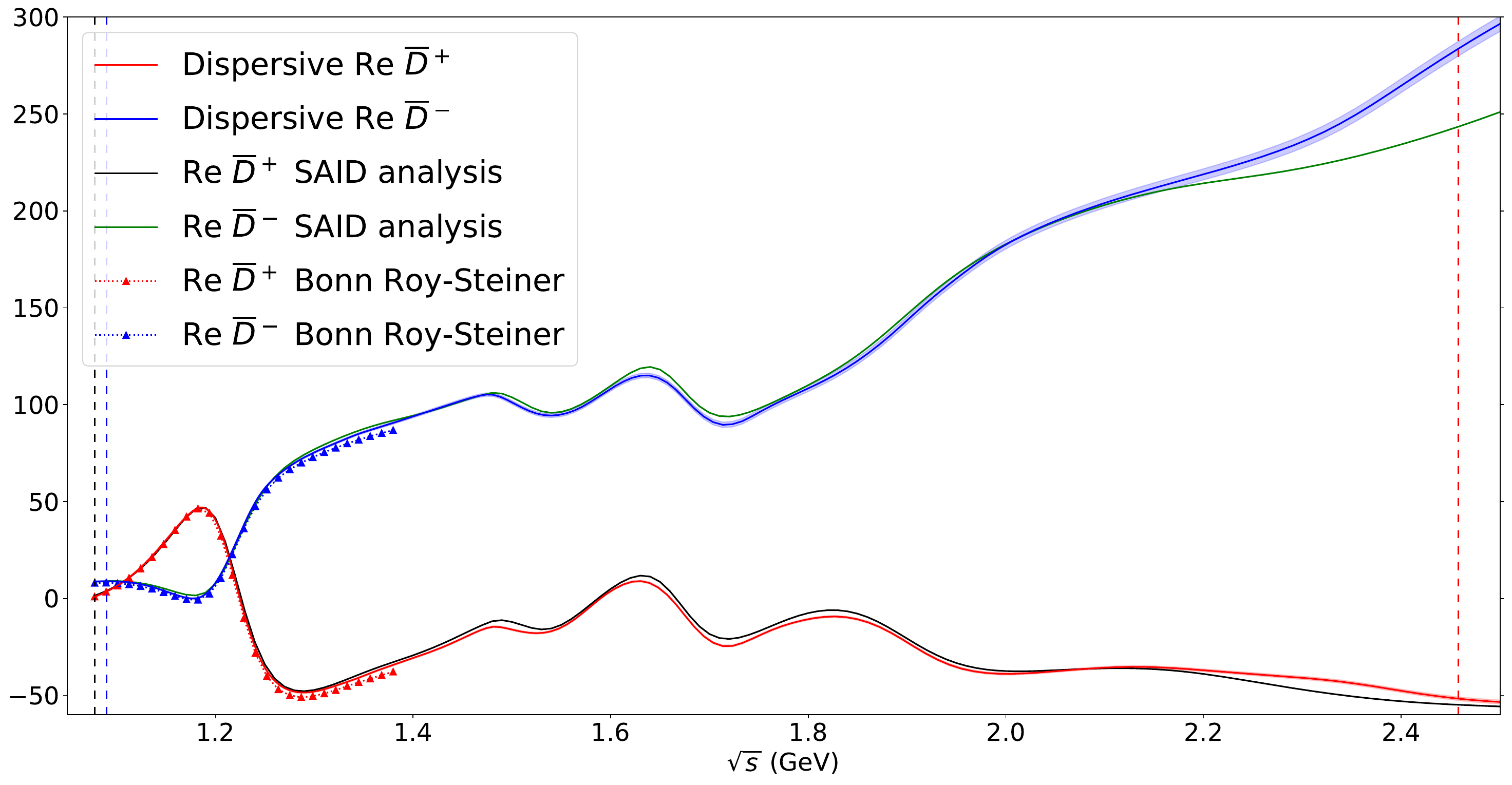}
\caption{Real parts of the $\bar{D}^\pm$ amplitudes from the forward
dispersion relations~\eqref{eq:FDR_+} and~\eqref{eq:FDR_-}, using the
fitted total cross-sections and Regge parametrization as input, compared with the Roy--Steiner analysis of~\cite{Hoferichter:2015hva} and the SAID/GWU partial-wave analysis~\cite{Workman:2012hx,SAID}.}
\label{fig:fdr}
\end{figure*}

The resulting $\bar{D}^\pm$ amplitudes are shown in Fig.~\ref{fig:fdr}, together with the Roy--Steiner solution of~\cite{Hoferichter:2015hva} and the SAID/GWU partial-wave analysis~\cite{Workman:2012hx,SAID}. At low energies, the dispersive reconstruction agrees with both analyses within uncertainties. In particular, the Roy--Steiner solution of~\cite{Hoferichter:2015hva} is fairly well reproduced even though we rely primarily on total cross-section data and only use Roy--Steiner input up to $k_{\rm lab}=64\,$MeV (the region between the first two vertical dashed lines). The agreement with the SAID/GWU solution is likewise very good up to 1.5 GeV, and fairly good up to $\sqrt{s}\simeq2.2\,\mathrm{GeV}$. Larger differences emerge only at higher energies, where the dispersive reconstruction becomes increasingly sensitive to the Regge input. Since the quoted upper limit of validity of the SAID/GWU solution is $\sqrt{s}\simeq2.46\,\mathrm{GeV}$, agreement beyond this energy is not expected; above $\sqrt{s}\simeq2.46\,\mathrm{GeV}$, the SAID/GWU solution departs markedly from both the experimental data and the present dispersive reconstruction.

\section{Results}\label{sec:results}

The analytic continuation of the forward amplitudes $D^\pm$ into the
complex-energy plane is performed using the $\mathrm{FDR}_{C_N}$ method of~\cite{Pelaez:2022qby,Pelaez:2025jrn}, which approximates an amplitude $F(s)$ by a continued fraction of order $N$,
\begin{equation}
    C_{N}(s)=F\left(s_{1}\right)\Big/\left(
    \vcenter{\hbox{$
    1+\frac{\displaystyle a_{1}(s-s_{1})}
    {\displaystyle 1+\frac{\displaystyle a_{2}(s-s_{2})}
    {\displaystyle \ddots\, a_{N-1}(s-s_{N-1})}}
    $}}
    \right),
    \label{eq:CN}
\end{equation}
where the coefficients $a_i$ are determined recursively from $C_N(s_i)=F(s_i)$ at a set of $N$ equally spaced sampling points $\{s_i\}$ in $\sqrt{s}$. In the present application, $F$ denotes any of the amplitudes $D^\pm$. In this way, pole positions and residues can be extracted from~\eqref{eq:CN} without imposing any relation between them, thus avoiding such a model dependence.

The systematic uncertainty associated with the analytic continuation
is estimated by varying the order $N$ from 15 to 51 (as explained in~\cite{Pelaez:2022qby,Pelaez:2025jrn}, the high-energy behavior of this amplitude calls for odd $N$), the sampling interval, and the parameters of the total cross-section parametrization. 
By studying such a large number of continued fractions with different orders, and therefore different number of parameters, any further model dependence is effectively included in our systematic uncertainties.
Poles that remain stable under these variations are retained, while spurious poles appearing only for particular choices of parameters are discarded following the criteria of~\cite{Pelaez:2022qby,Pelaez:2025jrn}. The spread of the stable pole determinations is included in the final uncertainty.

The isospin content of each pole is determined from the
combinations
\begin{equation}
    D^{1/2} = D^+ + 2D^-,
    \qquad
    D^{3/2} = D^+ - D^-,
\end{equation}
which project onto the $I=1/2$ and $I=3/2$ $\pi N$ channels, respectively. We find four stable poles in each isospin channel. As we 
advanced in the introduction, the lightest poles of the $I=3/2$ and $I=1/2$ channels are directly identifiable with the $\Delta(1232)$ and $N(1440)$ resonances, respectively.
However, this is not possible for the remaining six poles.  Nevertheless, these poles are not artifacts but arise as the combined effect of multiple resonances associated with the established $\Delta$- and $N$-spectrum. This is because amplitudes with different spin contribute to each FDR. As we will show in the next subsection, by using the SAID/GWU partial-wave parametrizations, the analytic continuation of FDRs cannot separate each resonance pole. Actually, multiple resonances seen in the partial-wave analysis combine into a single pole in the FDR analytic continuation, i.e., the number of poles observed from FDRs is less than that from partial waves.

Before turning to these results, a comment on the Riemann-sheet structure is in order. Since the amplitude is sampled on a segment on the real axis, the continued fraction provides the analytic continuation to the ``adjacent" or ``contiguous" sheet, which is the one obtained by continuously crossing all cuts opening at smaller energies than those in the segment. Let us recall that over each real segment between two consecutive thresholds, there is one specific adjacent sheet: between the first and second thresholds, it is the second sheet; between the second and third thresholds, it becomes the third sheet, and so on~\cite{Pelaez:2026fmu}. The poles reported here are therefore those of the sheet accessible from the sampling region, and we cannot make any claim about poles lying in other sheets, even if they are of interest~\cite{Doring:2025sgb,Doring:2025phq}. For the $N(1440)$, this is the same sheet in which the Roy--Steiner determination of~\cite{Hoferichter:2023ptl} was obtained, so that the two determinations refer to the same object and can be compared directly. We stress that this is not a limitation specific to continued fractions, but a general feature of any continuation performed from the physical region.

A related caveat concerns sampling intervals that encompass a threshold. Should the poles contiguous from below and from above be well separated, a single continued fraction sampled across the threshold would be forced to approximate both analytic structures at once, and could return an intermediate result with an ambiguous sheet interpretation, and without any warning in the output that this has occurred.  
Multi-pion $n\pi N$ thresholds, which open at every pion mass, are unavoidable in any interval; their opening is suppressed by increasing powers of the phase space with $n$, they induce no visible structure in the total cross-sections; and the associated discontinuities are therefore negligible. Some of our intervals also encompass two-body thresholds (see the intervals given in Sect.~\ref{sec:poles}), most notably $\eta N$ and those involving strange particles, which could not be avoided while retaining a stable extraction. In such cases, the
sheet reached is the contiguous one to the segment as a whole, and the poles we quote should be understood accordingly. In practice, this would be reflected in the systematic uncertainties, since the extracted pole position would drift as the sampling segment is varied; we find central values stable under such variations, with only the uncertainties growing away from the
optimal segment.

Returning to the structure of this section, in a second subsection, we will show that the resolution of resonance poles deteriorates when the forward amplitudes are obtained from total cross-section data instead of being reconstructed from partial-wave parametrizations. 
This is because real data is a discrete set of points and carries uncertainty.
Still, we will nevertheless find that the poles found from the FDRs and total cross-section data have physical content and can be related to the $\Delta$ and $N$ spectra. This we will do by writing sum rules relating their residues.
In the last subsection covering our results, we present a model-independent determination of the $\Delta^{0}-\Delta^{++}$ difference from FDRs and total cross-section data.

\subsection{FDR poles from partial-wave parametrizations}\label{sec:validation}

The $\mathrm{FDR}_{C_N}$ method has previously been applied to
meson spectroscopy, where it was shown to reproduce the resonance
content of dispersive $\pi\pi$ scattering amplitudes~\cite{Pelaez:2022qby,Pelaez:2025jrn}. In the present application to $\pi N$, the situation is considerably more challenging, since several resonances with different spin quantum numbers may contribute simultaneously to the same forward amplitude and cannot be resolved individually.


\begin{table*}[t]
\footnotesize
\renewcommand{\arraystretch}{1.5}
\begin{threeparttable}
\begin{tabular}{ccl}
\hline\hline
\multicolumn{3}{c}{$I=\frac{1}{2}$ channel} \\ \hline
Associated SAID/GWU resonance(s)
  & $\mathrm{FDR}_{C_N}$ pole (MeV)
  & SAID/GWU pole (MeV)~\cite{Arndt:2006bf} \\ \hline
$N(1440)\,\frac{1}{2}^+$
  & $(1365\pm3)-i\,(79\pm4)$
  & $1359-i\,81$ \\ \hline
$N(1520)\,\frac{3}{2}^-$
  & \multirow{2}{*}{$(1513.0\pm2.4)-i\,(55.4\pm1.9)$}
  & $1515-i\,56.5$ \\
$N(1535)\,\frac{1}{2}^-$
  & & $1502-i\,47.5$ \\ \hline
$N(1650)\,\frac{1}{2}^-$
  & \multirow{4}{*}{$(1670.6\pm1.5)-i\,(59.5\pm1.3)$}
  & $1648-i\,40$ \\
$N(1675)\,\frac{5}{2}^-$ & & $1657-i\,69.5$ \\
$N(1680)\,\frac{5}{2}^+$ & & $1674-i\,57.5$ \\
$N(1720)\,\frac{3}{2}^+$ & & $1666-i\,177.5$ \\ \hline
$N(1860)\,\frac{5}{2}^+$
  & $(1786\pm3)-i\,(116\pm3)$
  & $1785-i\,122$\,\tnote{(a)} \\ \hline
$N(2190)\,\frac{7}{2}^-$
  & \multirow{4}{*}{$(2131\pm12)-i\,(301\pm11)$}
  & $2070-i\,260$ \\
$N(2220)\,\frac{9}{2}^+$  & & $2199-i\,186$ \\
$N(2245)\,\frac{11}{2}^+$ & & $2203-i\,66.5$ \\
$N(2250)\,\frac{9}{2}^-$  & & $2217-i\,215.5$ \\ \hline\hline
\end{tabular}
\begin{tablenotes}
\footnotesize
\item[(a)] For the $N(1860)$, Ref.~\cite{Arndt:2006bf} quotes the pole position $(1807-i\,54.5)\,\mathrm{MeV}$, whereas the current SAID/GWU database yields $(1785-i\,122)\,\mathrm{MeV}$, corresponding to the solution analyzed here.
\end{tablenotes}
\end{threeparttable}
\caption{Poles extracted from the analytic continuation of $D^{1/2}$ reconstructed from the SAID/GWU partial-wave solution~\cite{Workman:2012hx,SAID}. The corresponding SAID/GWU pole positions for the resonance assignments shown in the first column are listed for comparison.}
\label{tab:SAID_N}
\end{table*}

\begin{table*}[t]
\footnotesize
\renewcommand{\arraystretch}{1.5}
\begin{tabular}{ccc}
\hline\hline
\multicolumn{3}{c}{$I=\frac{3}{2}$ channel} \\ \hline
Associated SAID/GWU resonance(s)
  & $\mathrm{FDR}_{C_N}$ pole (MeV)
  & SAID/GWU pole (MeV)~\cite{Arndt:2006bf} \\ \hline
$\Delta(1232)\,\frac{3}{2}^+$
  & $(1211.5\pm0.1)-i\,(50.2\pm0.1)$
  & $1211-i\,49.5$ \\ \hline
$\Delta(1600)\,\frac{3}{2}^+$
  & $(1431\pm21)-i\,(192\pm11)$
  & $1457-i\,200$ \\ \hline
$\Delta(1620)\,\frac{1}{2}^-$
  & $(1595.3\pm1.4)-i\,(66.7\pm1.2)$
  & $1595-i\,67.5$ \\ \hline
$\Delta(1700)\,\frac{3}{2}^-$
  & $(1641\pm14)-i\,(145\pm12)$
  & $1632-i\,126.5$ \\ \hline
$\Delta(1905)\,\frac{5}{2}^+$
  & \multirow{2}{*}{$(1806\pm15)-i\,(134\pm14)$}
  & $1819-i\,123.5$ \\
$\Delta(1910)\,\frac{1}{2}^+$
  & & $1771-i\,239.5$ \\ \hline
$\Delta(1930)\,\frac{5}{2}^-$
  & \multirow{2}{*}{$(1883.2\pm2.0)-i\,(115\pm3)$}
  & $2001-i\,193.5$ \\
$\Delta(1950)\,\frac{7}{2}^+$
  & & $1876-i\,113.5$ \\ \hline
$\Delta(2420)\,\frac{11}{2}^+$
  & $(2353\pm18)-i\,(269\pm15)$
  & $2529-i\,310.5$ \\ \hline\hline
\end{tabular}
\caption{Poles extracted from the analytic continuation of $D^{3/2}$ reconstructed from the SAID/GWU partial-wave solution~\cite{Workman:2012hx,SAID}. The corresponding SAID/GWU pole positions~\cite{Arndt:2006bf} for the resonance assignments shown in the first column are listed for comparison.}
\label{tab:SAID_Delta}
\end{table*}

A useful benchmark is provided by the SAID/GWU solution~\cite{Workman:2012hx,SAID}. Starting from the SAID/GWU partial waves, we reconstruct the corresponding forward $D^I$ amplitudes and apply the same analytic-continuation procedure used throughout this work. Since the resonance poles of the underlying partial-wave solution are known independently, this provides a direct test of the ability of the $\mathrm{FDR}_{C_N}$ method to recover the resonance content encoded in the FDRs in an ideal situation, i.e., without uncertainties and the need for a fit to discrete data points.

The resulting poles are collected in Tables~\ref{tab:SAID_N} and~\ref{tab:SAID_Delta}. Each $\mathrm{FDR}_{C_N}$ pole is compared with the SAID/GWU pole positions corresponding to the resonance assignments of the underlying partial-wave analysis. Since the SAID/GWU analysis does not provide statistical uncertainties, no uncertainties are quoted in the tables for their pole positions.

As expected, the FDR$_{C_{N}}$ method does not resolve resonances that are close in energy and share the same isospin quantum numbers; instead, it identifies stable poles corresponding to the combined effect of nearby resonances in each energy region. This limitation is particularly evident in the $I=1/2$ channel, where several SAID/GWU poles cluster within relatively narrow energy intervals, whereas in the $I=3/2$ sector, most poles remain associated with one or at most two nearby resonances.

The extracted poles are in good agreement with the corresponding SAID/GWU pole positions throughout the energy range, most strikingly for the $\Delta(1232)$ and Roper resonances, whose pole positions are reproduced almost exactly. This agreement extends to the higher-mass regions, where several resonances overlap. The only notable exception is the heaviest dispersive pole with $I=3/2$, but the pole of the SAID/GWU analysis lies beyond its strict validity region. 
Note that in this case, the $\mathrm{FDR}_{C_N}$ pole position is nevertheless in good agreement with the current RPP estimate~\cite{ParticleDataGroup:2026aaa}.

This benchmark demonstrates that the $\mathrm{FDR}_{C_N}$ method reliably reconstructs the overall singularity structure as encoded in the FDRs, even in the presence of substantial resonance overlap. It therefore provides a non-trivial validation of the analytic-continuation procedure in the baryon sector and supports the interpretation of the poles extracted from the experimental total cross-section data presented in the following subsections.

\subsection{Pole parameters and residue sum rules}
\label{sec:poles}

The analytic continuation of the amplitudes $D^{1/2}$ and $D^{3/2}$ reveals four stable poles in each isospin sector. Their pole positions $\sqrt{s_{\mathrm{pole}}}$, residue moduli $|R|$, and residue phases $\phi$ are collected in Tables~\ref{tab:N_res} and~\ref{tab:Delta_res}. As demonstrated in Sect.~\ref{sec:validation}, the $\mathrm{FDR}_{C_N}$ method reliably recovers the dominant poles of the forward amplitudes, but does not resolve resonances that are close in energy and share the same isospin quantum numbers. The same pattern is observed here: the lightest pole in each sector can be associated unambiguously with a single resonance, namely the Roper resonance $N(1440)$ in the $I=1/2$ channel and the $\Delta(1232)$ in the $I=3/2$ channel, while the remaining poles receive contributions from several nearby resonances, reflecting both the limited resolving power of forward amplitudes and the overlap of resonance signals in the total cross-sections.

\begin{table}[h]
\footnotesize
\resizebox{.46\textwidth}{!}{
\renewcommand{\arraystretch}{1.6}
\begin{tabular}{cccc}
\hline\hline
\multicolumn{4}{c}{$I=\frac{1}{2}$ determinations} \\ \hline
Resonances $J^P$
  & $\sqrt{s_{\mathrm{pole}}}$ (MeV)
  & $|R|$ (GeV)
  & $\phi$ ($^\circ$) \\ \hline
$N(1440)$ $\frac{1}{2}^+$
  & $(1334^{+21}_{-26})-i(81^{+25}_{-21})$
  & $6^{+4}_{-3}$
  & $29^{+43}_{-46}$ \\ \hline
$N(1520)$ $\frac{3}{2}^-$
  & \multirow{2}{*}{$(1520^{+20}_{-19})-i(65^{+16}_{-15})$}
  & \multirow{2}{*}{$18^{+9}_{-8}$}
  & \multirow{2}{*}{$197^{+33}_{-34}$} \\
$N(1535)$ $\frac{1}{2}^-$
  & & & \\ \hline
$N(1650)$ $\frac{1}{2}^-$
  & \multirow{6}{*}{$(1668\pm14)-i(62^{+10}_{-11})$}
  & \multirow{6}{*}{$32\pm13$}
  & \multirow{6}{*}{$172^{+26}_{-22}$} \\
$N(1675)$ $\frac{5}{2}^-$ & & & \\
$N(1680)$ $\frac{5}{2}^+$ & & & \\
$N(1700)$ $\frac{3}{2}^-$ & & & \\
$N(1710)$ $\frac{1}{2}^+$ & & & \\
$N(1720)$ $\frac{3}{2}^+$ & & & \\ \hline
$N(2060)$ $\frac{5}{2}^-$
  & \multirow{6}{*}{$(2117^{+40}_{-42})-i(260^{+51}_{-50})$}
  & \multirow{6}{*}{$133^{+51}_{-47}$}
  & \multirow{6}{*}{$137\pm31$} \\
$N(2100)$ $\frac{1}{2}^+$ & & & \\
$N(2120)$ $\frac{3}{2}^-$ & & & \\
$N(2190)$ $\frac{7}{2}^-$ & & & \\
$N(2220)$ $\frac{9}{2}^+$ & & & \\
$N(2250)$ $\frac{9}{2}^-$ & & & \\ \hline\hline
\end{tabular}}
\caption{Pole parameters extracted from the analytic continuation of
$D^{1/2}$. Except for the $N(1440)$, each structure receives contributions from more than one resonance.}
\label{tab:N_res}
\end{table}

\begin{table}[h]
\footnotesize
\resizebox{.46\textwidth}{!}{
\renewcommand{\arraystretch}{1.6}
\begin{tabular}{cccc}
\hline\hline
\multicolumn{4}{c}{$I=\frac{3}{2}$ determinations} \\ \hline
Resonances $J^P$
  & $\sqrt{s_{\mathrm{pole}}}$ (MeV)
  & $|R|$ (GeV)
  & $\phi$ ($^\circ$) \\ \hline
$\Delta(1232)$ $\frac{3}{2}^+$
  & $(1211\pm4)-i(52\pm4)$
  & $25\pm10$
  & $157\pm18$ \\ \hline
$\Delta(1620)$ $\frac{1}{2}^-$
  & \multirow{2}{*}{$(1622\pm20)-i(88\pm21)$}
  & \multirow{2}{*}{$10\pm4$}
  & \multirow{2}{*}{$153\pm37$} \\
$\Delta(1700)$ $\frac{3}{2}^-$
  & & & \\ \hline
$\Delta(1900)$ $\frac{1}{2}^-$
  & \multirow{6}{*}{$(1888\pm8)-i(117\pm8)$}
  & \multirow{6}{*}{$32\pm6$}
  & \multirow{6}{*}{$157\pm10$} \\
$\Delta(1905)$ $\frac{5}{2}^+$ & & & \\
$\Delta(1910)$ $\frac{1}{2}^+$ & & & \\
$\Delta(1920)$ $\frac{3}{2}^+$ & & & \\
$\Delta(1930)$ $\frac{5}{2}^-$ & & & \\
$\Delta(1950)$ $\frac{7}{2}^+$ & & & \\ \hline
$\Delta(2300)$ $\frac{9}{2}^+$
  & \multirow{3}{*}{$(2355\pm15)-i(133^{+13}_{-12})$}
  & \multirow{3}{*}{$14\pm3$}
  & \multirow{3}{*}{$131^{+17}_{-18}$} \\
 $\Delta(2400)$ $\frac{9}{2}^-$
  & & & \\
$\Delta(2420)$ $\frac{11}{2}^+$
  & & & \\ \hline\hline
\end{tabular}}
\caption{Pole parameters extracted from the analytic continuation of
$D^{3/2}$. Except for the $\Delta(1232)$, each structure receives contributions from more than one resonance.}
\label{tab:Delta_res}
\end{table}

\begin{figure*}[ht]
\centering
\includegraphics[width=\textwidth]{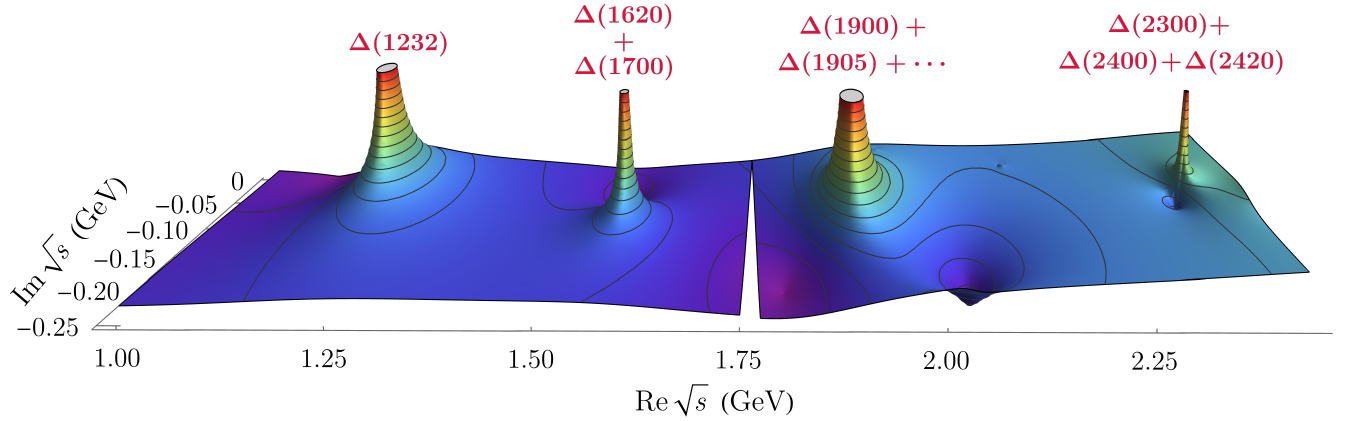}
\caption{Representative analytic continuation of $D^{3/2}$, displayed over
the ranges $[1.15,1.75]\,\mathrm{GeV}$ and $[1.75,2.5]\,\mathrm{GeV}$, for a
fixed order $N$. Note that these are the ranges over which the continuation
is shown, not the real-axis intervals from which the amplitude is sampled to construct the continued fraction (see text). The four stable $I=3/2$ poles are indicated.
}
\label{fig:Deltas_rep}
\end{figure*}

\begin{figure*}[ht]
\centering
\includegraphics[width=\textwidth]{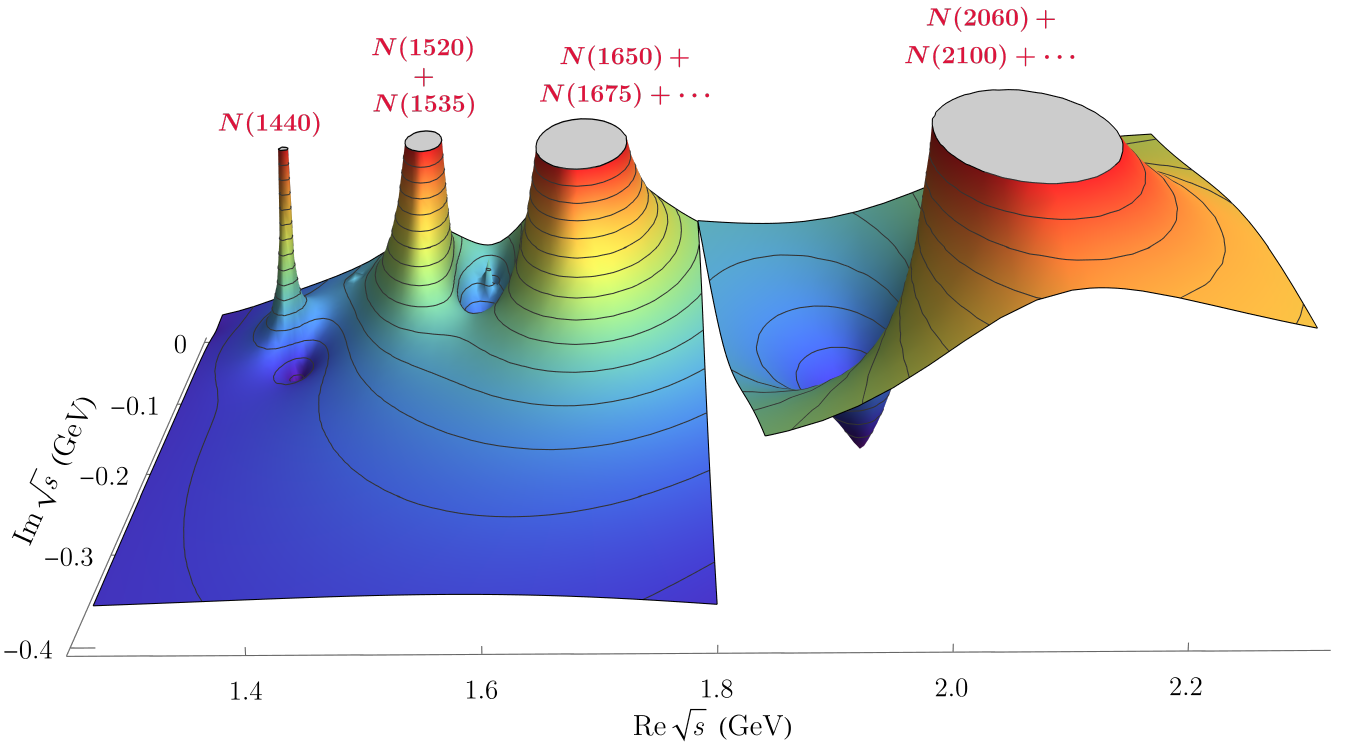}
\caption{Representative analytic continuation of $D^{1/2}$, displayed over
the ranges $[1.25,1.8]\,\mathrm{GeV}$ and $[1.8,2.3]\,\mathrm{GeV}$, for a
fixed order $N$. Note that these are the ranges over which the continuation
is shown, not the real-axis intervals from which the amplitude is sampled to construct the continued fraction (see text). The four stable $I=1/2$ poles are indicated.
}
\label{fig:N_rep}
\end{figure*}

Representative analytic continuations of $D^{1/2}$ and $D^{3/2}$ are shown in Figs.~\ref{fig:Deltas_rep} and~\ref{fig:N_rep}, which are displayed over the ranges given in the corresponding captions. These should not be confused with the real-axis intervals from which the amplitude is sampled to construct the continued fraction. 
The latter are selected by the stability of the resulting pole and, where possible, within regions free of relevant inelastic thresholds. For the $N(1440)$, for instance, we sample $[1.25,1.45]\,\mathrm{GeV}$, which lies
between the $\pi\pi N$ and $\eta N$ thresholds, whereas for the remaining
$I=1/2$ poles we use $[1.45,1.65]$, $[1.6,1.8]$, and $[2.05,2.3]\,\mathrm{GeV}$. The corresponding intervals for the
$I=3/2$ channel are $[1.15,1.35]$, $[1.55,1.75]$, $[1.75,1.95]$, and $[2.3,2.5]\,\mathrm{GeV}$.

\begin{figure}[h]
\centering
\includegraphics[width=0.47\textwidth]{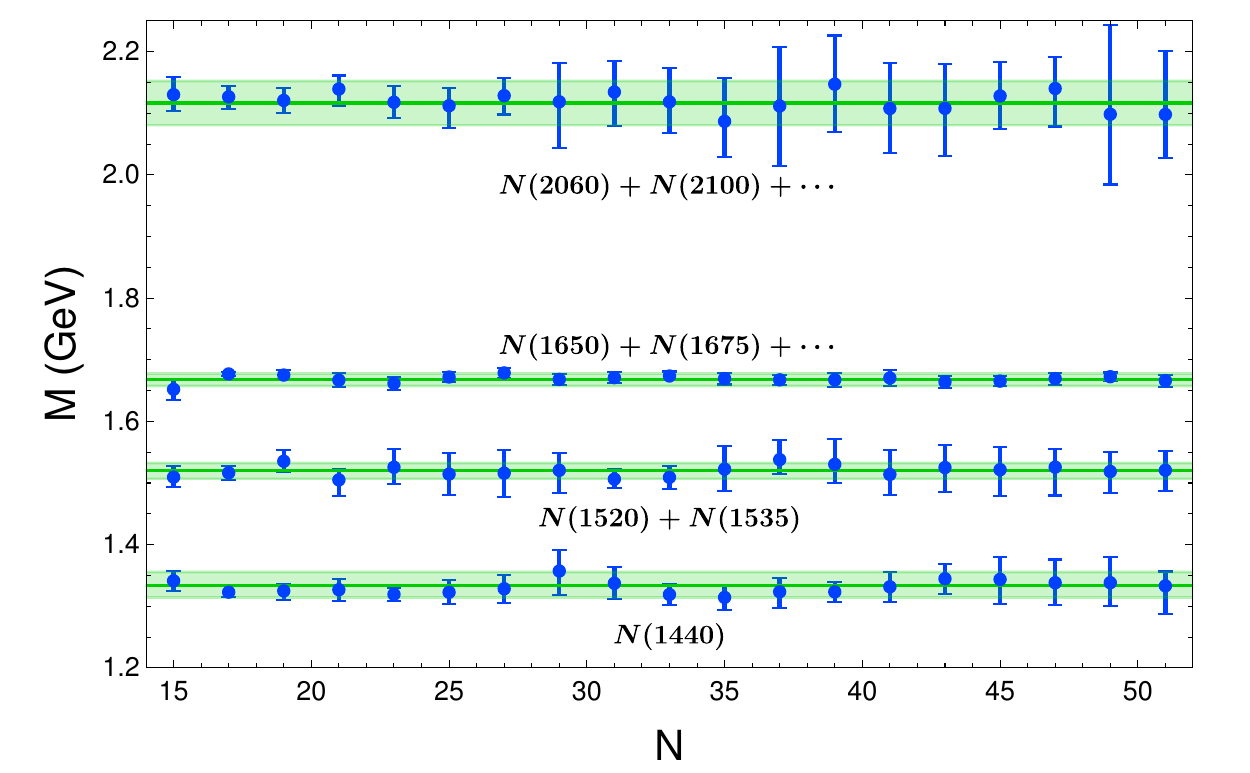}
\caption{Pole masses of the $I=1/2$ structures as a function of the
continued-fraction order $N$, illustrating the stability of the
extraction.}
\label{fig:N_N}
\end{figure}

\begin{figure}[h]
\centering
\includegraphics[width=0.47\textwidth]{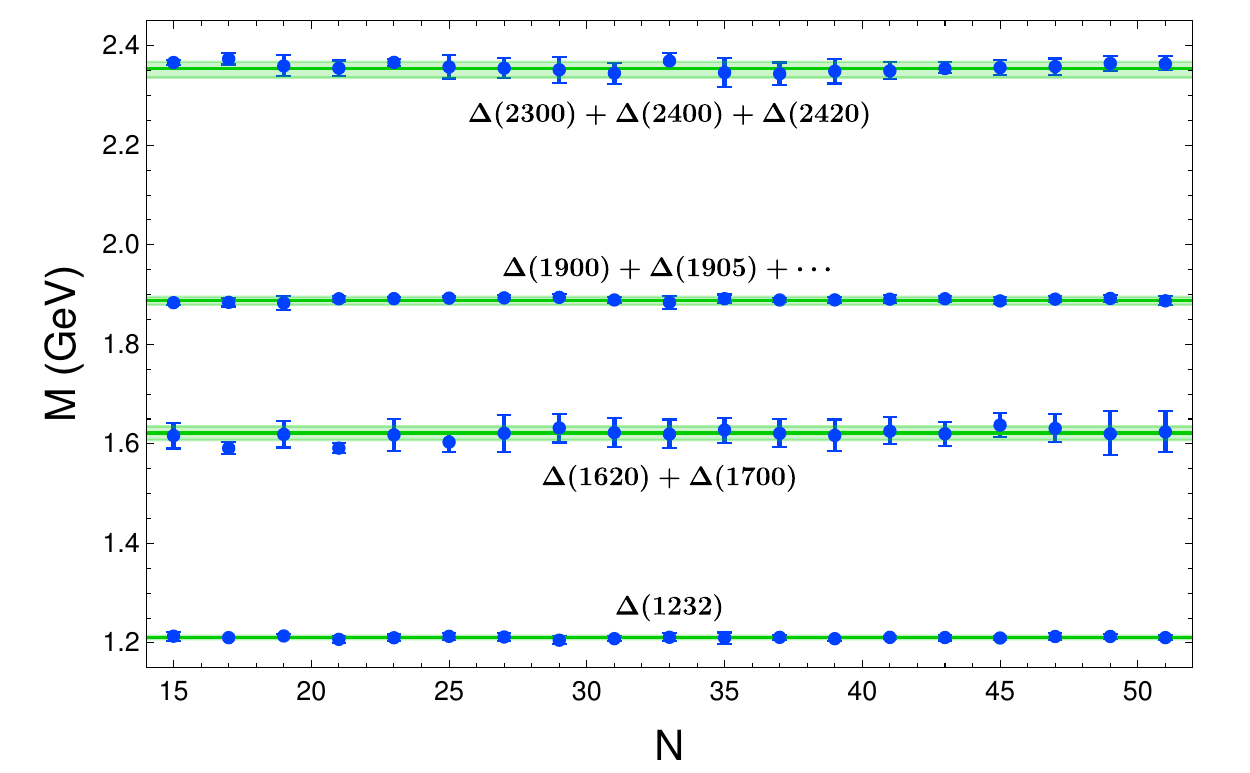}
\caption{Pole masses of the $I=3/2$ structures as a function of the
continued-fraction order $N$, illustrating the stability of the
extraction.}
\label{fig:Deltas_N}
\end{figure}

The stability of the extracted poles under variation of the continued-fraction order $N$ is illustrated in Figs.~\ref{fig:N_N} and~\ref{fig:Deltas_N}; the green bands indicate the uncertainties quoted in the tables. 

The extracted $\Delta(1232)$ pole is in excellent agreement with both the current RPP estimate~\cite{ParticleDataGroup:2026aaa}, $\sqrt{s_{\rm pole}}=(1209\text{--}1211) -i(49\text{--}51)$\,MeV, and the recent Roy--Steiner determination of~\cite{Hoferichter:2023mgy}, $(1209.5\pm1.1)-i(49.2\pm0.6)$\,MeV, providing a useful consistency check of the present reconstruction.
Of particular interest is the pole associated with the Roper resonance, since its effect is imperceptible in the total cross-section data shown in Fig.~\ref{fig:cross-sections}. Still, the analytic continuation of the $I=1/2$ amplitude yields a stable pole in the corresponding energy region. The extracted pole is compatible within uncertainties with the current RPP estimate~\cite{ParticleDataGroup:2026aaa}, $\sqrt{s_{\rm pole}}=(1360\text{--}1380) -i(90\text{--}102)$\,MeV
and shows only a mild tension with the recent Roy--Steiner determination of~\cite{Hoferichter:2023mgy}, $(1374\pm5)-i(107\pm10)$\,MeV.

For the heavier poles, several resonances lie close in energy and cannot be resolved individually from the forward amplitudes alone. The assignment of a group of RPP resonances in Tables~\ref{tab:N_res} and~\ref{tab:Delta_res} to each of the dispersive FDR$_{C_N}$ poles is first made by their proximity. This assignment is confirmed next by using sum rules relating their residues, but only if their contribution to the sum rules is not negligible.

The FDR$_{C_N}$ analytic continuation determines a unique complex residue $R^{\rm FDR}$ for each dispersive pole, obtained by encircling it with a closed contour and applying Cauchy's Theorem. 
If the contour is chosen to also enclose the
group of resonance poles believed to generate the dispersive pole, their residues $R_j$ give rise to the following sum rule:
\begin{equation}
    R^{\rm FDR}=\sum_j R_j,
\end{equation}
where $j$ stands for the $j$-th resonance pole assigned to the dispersive pole.
Thus, if our assignments are correct, our dispersive residues in Tables~\ref{tab:N_res} and~\ref{tab:Delta_res} should be compatible with the sum of the residues of their assigned RPP resonances. 
Since the RPP adopts a different normalization for the residue, denoted by $r^{\text{RPP}}$, a conversion factor must be applied first:
\begin{equation}
    R^{\text{RPP}}_{j} =K^\text{RPP}_{j}|r_{j}^\text{RPP}|  e^{i\displaystyle \phi_{j}^\text{RPP}} ,
\end{equation}
where we have defined the following kinematic factor
\begin{equation}
    K^{\text{RPP}}_{j} =-\frac{8\pi \sqrt{s^\text{RPP}_{j}}k_\text{lab}(s^\text{RPP}_{j})(J_{j}+\frac{1}{2})}{q_\text{cm}^2(s^\text{RPP}_{j})},
\end{equation}
with $J_{j}$ the total angular momentum of the $j$-th resonance.

Table~\ref{tab:residue_sum_rules} compares the complex residues extracted from the analytic continuation with those obtained by combining the RPP pole residues of the corresponding resonances~\cite{ParticleDataGroup:2026aaa}. For the $I=1/2$ channel, the comparison includes all resonances assigned to each pole in Table~\ref{tab:N_res}. For the $I=3/2$ channel, where several alternative assignments are possible, we list the combinations favored by the simultaneous comparison of the residue moduli and phases.


\onecolumngrid

\vspace{2mm}

\begin{table*}[h]
\centering
\small
\renewcommand{\arraystretch}{1.15}
\setlength{\tabcolsep}{2mm}
\begin{tabular}{lcccc}
\hline\hline
Pole
  & $|R^{\mathrm{FDR}}|$ (GeV)
  & $|R^{\mathrm{RPP}}| $ (GeV)  & $\phi^{\mathrm{FDR}}$ ($^\circ$)
  & $\phi^{\mathrm{RPP}}$ ($^\circ$) \\
\hline
$N(1440)$
  & $6^{+4}_{-3}$     & $7.8\pm0.7$   & $29^{+43}_{-46}$  & $94\pm10$ \\
$N(1520)+N(1535)$
  & $18^{+9}_{-8}$    & $12.8\pm1.6$  & $197^{+33}_{-34}$ & $168\pm7$ \\
$N(1650) + \cdots + N(1720)$
  & $32\pm13$         & $34\pm6$     & $172^{+26}_{-22}$ & $136\pm10$   \\
$N(2060) + \cdots + N(2250)$
  & $133^{+51}_{-47}$ & $77\pm20$   & $137\pm31$        & $143\pm15$    \\
\hline
$\Delta(1232)$
  & $25\pm10$         & $18.3\pm0.7$  & $157\pm18$        & $142.1\pm2.1$ \\
$\Delta(1620)+\Delta(1700)$
  & $10\pm4$          & $7\pm4$      & $153\pm37$        & $144\pm25$   \\
$\Delta(1905)+\Delta(1910)+\Delta(1950)$
  & $32\pm6$          & $36\pm7$   & $157\pm10$        & $133\pm11$     \\
$\Delta(2420)$
  & $14\pm3$          & $17\pm8$    & $131^{+17}_{-18}$ & $157\pm36$    \\
\hline\hline
\end{tabular}
\caption{Comparison of the residue moduli $|R|$ and phases $\phi$ extracted from the analytic continuation of the FDRs with those obtained from the combined RPP pole residues for the resonance assignments in Tables~\ref{tab:N_res} and~\ref{tab:Delta_res}.}
\label{tab:residue_sum_rules}
\end{table*}

\vspace{2mm}

\twocolumngrid


In the $I=1/2$ channel, the agreement between the dispersive residues and the combined RPP contributions is generally satisfactory, both in modulus and phase. The pole at $\sqrt{s_{\mathrm{pole}}}\simeq(1520-i\,65)\,\mathrm{MeV}$ is well accounted for by the $N(1520)$ and $N(1535)$. Likewise, the pole at $\sqrt{s_{\mathrm{pole}}}\simeq(1668-i\,62)\,\mathrm{MeV}$ is reproduced remarkably well by the established resonances between 1650 and 1720~MeV, with the dominant contributions arising from the $N(1650)$, $N(1675)$, and $N(1680)$. Additional contributions from the $N(1700)$, $N(1710)$, and $N(1720)$ remain compatible with the dispersive determination, although the agreement obtained already from the lightest states indicates that they are not required to account for the observed pole. 

For the heaviest pole, the residue is dominated by the $N(2190)$, $N(2220)$, and $N(2250)$ resonances, with additional contributions from the $N(2060)$, $N(2100)$, and especially the $N(2120)$, improving the agreement with the dispersive residue modulus. The combined RPP residue modulus remains somewhat smaller than the dispersive determination, while the residue phase is in excellent agreement. An even better agreement is obtained when using the larger residue values for the $N(2190)$ and $N(2220)$ reported in~\cite{Anisovich:2011fc,Svarc:2014aga}.

In the $I=3/2$ channel, the pole at $\sqrt{s_{\mathrm{pole}}}\simeq(1622-i\,88)\,\mathrm{MeV}$ is consistent with the combined contribution of the $\Delta(1620)$ and $\Delta(1700)$ resonances. While the residue modulus alone does not fully discriminate among nearby assignments, the residue phase provides an additional constraint and strongly favors the $\Delta(1620)+\Delta(1700)$ interpretation over alternatives involving the $\Delta(1600)$. 

The pole at $\sqrt{s_{\mathrm{pole}}}\simeq(1888-i\,117)\,\mathrm{MeV}$ is
largely dominated by the $\Delta(1950)$, with additional contributions from the $\Delta(1905)$ and $\Delta(1910)$. The agreement in both modulus and phase strongly favors this assignment, while the inclusion of additional nearby resonances deteriorates the agreement with the dispersive residue.

Finally, the heaviest pole is compatible with a dominant contribution from the $\Delta(2420)$. The inclusion of additional nearby resonances deteriorates the agreement with the dispersive residue, while the comparison favors the smaller residue values obtained in~\cite{Cutkosky:1979fy}.

The pole positions and residues consistently support the resonance assignments in Tables~\ref{tab:N_res} and~\ref{tab:Delta_res}. While they do not allow a unique decomposition of each pole into individual resonances, they provide non-trivial constraints on the dominant resonance content of each energy region and establish a quantitative connection between the dispersive reconstruction and the established $\pi N$ resonance spectrum.

\subsection{\texorpdfstring{$\Delta(1232)$}{Delta(1232)} isospin-breaking effects}\label{sec:IB_corr}

Electromagnetic interactions and the up--down quark-mass difference induce small but non-vanishing differences between the pole parameters of the $\Delta^0$ and $\Delta^{++}$ charged states, making the $\Delta(1232)$ region a sensitive probe of isospin-breaking effects in $\pi N$ scattering. To retain sensitivity to these effects, the analysis in this subsection is performed directly on the physical $\pi^\pm p$ total cross-sections, without applying the corrections discussed in Sect.~\ref{sec:corrections}. In principle, electromagnetic interactions introduce additional singularities not included in the dispersive framework, so that the resulting amplitudes no longer satisfy the forward dispersion relations exactly.
Since the $\Delta^0-\Delta^{++}$ splitting is extracted from the pole associated with the dominant right-hand-cut contribution, however, the neglected electromagnetic singularities are expected to generate only subleading effects. The procedure therefore provides a meaningful estimate of the $\Delta^0-\Delta^{++}$ splitting.

The forward amplitudes $D_+\equiv D(\pi^+p)$ and $D_-\equiv D(\pi^-p)$ in~\eqref{eq:amplitude_relations} are sensitive to the two charged states in complementary ways. For $D_+$, the dominant $\Delta(1232)$ contributes through the physical right-hand cut and is primarily sensitive to the $\Delta^{++}$ resonance, while the $\Delta^0$ contributes only through the crossed $u$ channel. The situation is reversed for $D_-$. Comparing the pole parameters extracted from the analytic continuations of $D_+$ and $D_-$ thus provides sensitivity to the differences between the two charged states.
Nevertheless, since the physical amplitudes $D_+$ and $D_-$ are normalized differently in the isospin basis, the corresponding pole residues cannot be compared directly. In the isospin limit, $D_+=D^{3/2}$, whereas $D_- =\frac{1}{3}D^{3/2}+\frac{2}{3}D^{1/2}$. Near the $\Delta(1232)$ resonance, where the $I=3/2$ contribution dominates, the residue associated with the $\Delta^0$ in $D_-$ is therefore expected to be smaller than that of the $\Delta^{++}$ in $D_+$ by a factor of three. We account for this normalization difference by comparing $|R_{\Delta^0}|$ with $|R_{\Delta^{++}}|/3$.

\begin{figure}[h]
\centering
\includegraphics[width=0.47\textwidth]{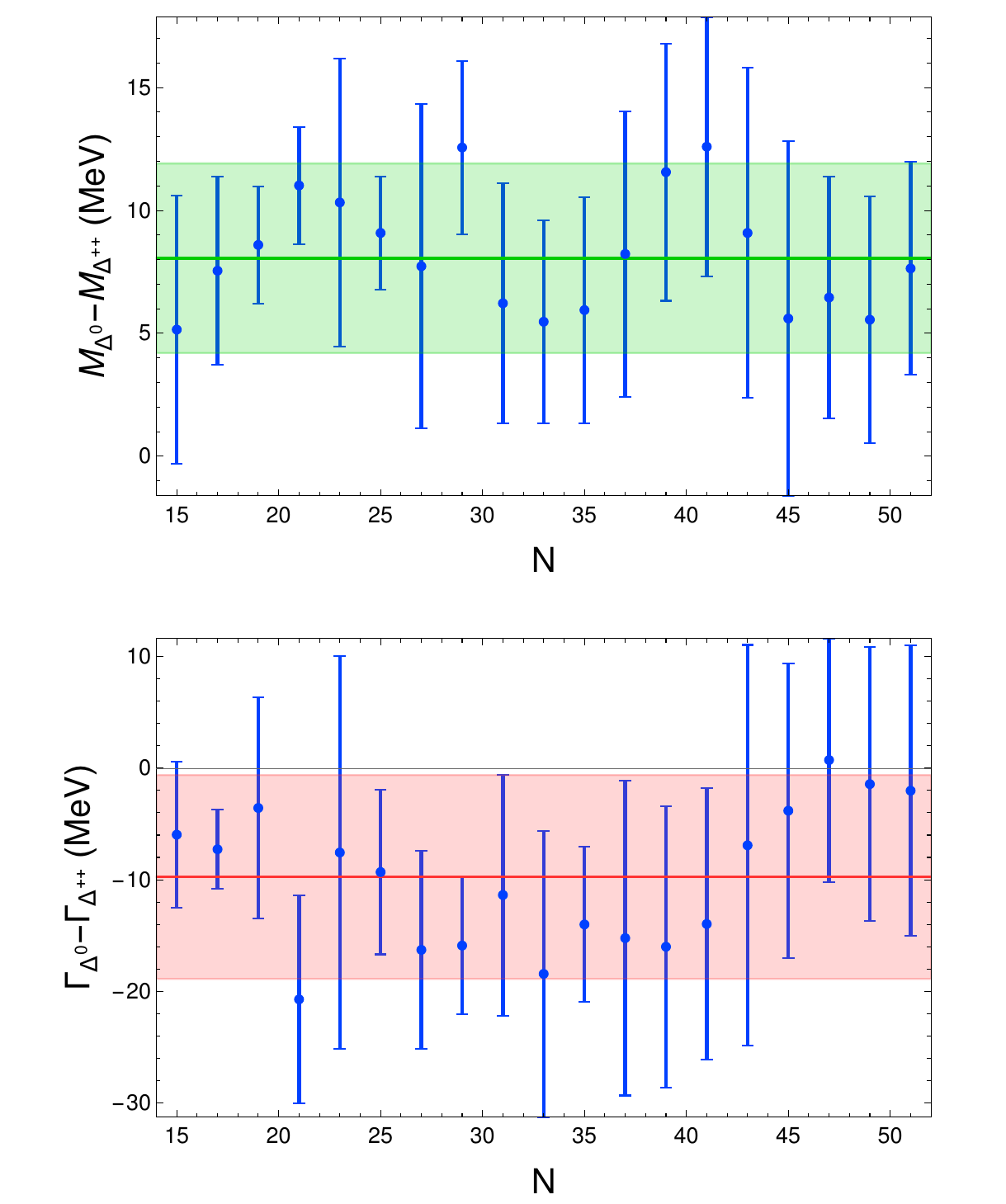}
\caption{Differences between the pole parameters extracted from the analytic continuations of $D_+$ and $D_-$ in the $\Delta(1232)$
region, shown as a function of the continued-fraction order $N$. The horizontal bands indicate the final averages and their uncertainties.}
\label{fig:Delta-IV}
\end{figure}

The mass and width differences as a function of the order $N$ of the continued fraction are shown in Fig.~\ref{fig:Delta-IV}, with uncertainties estimated following the same procedure as in Sect.~\ref{sec:results}. Averaging over the range of $N$ we have considered, we find
\begin{equation}
\begin{aligned}
M_{\Delta^0}-M_{\Delta^{++}}
  &= (8.1\pm3.9)\,\mathrm{MeV}, \\
\Gamma_{\Delta^0}-\Gamma_{\Delta^{++}}
  &= (-9.8\pm9.2)\,\mathrm{MeV}, \\
|R_{\Delta^0}|-|R_{\Delta^{++}}|/3
  &= (-0.8\pm1.2)\,\mathrm{GeV},\\
  |\phi_{\Delta^0}|-|\phi_{\Delta^{++}}|
  &= (16\pm14)\,^\circ.
\end{aligned}
\end{equation}
Our result favors a heavier $\Delta^0$, in agreement with the previous dedicated studies ~\cite{Pedroni:1978it,Abaev:1995cx,Bernicha:1995gg,Gridnev:2004mk} listed in the RPP, which provide the Breit-Wigner mass difference between 2.2 and 2.9 MeV with individual uncertainties of $0.3$ to 0.7 MeV. 

Quantitatively, however, our determination is much less precise, with an uncertainty of about $50\%$, so the difference with the Breit--Wigner values is not statistically significant. The present extraction is based directly on the pole structure of the forward amplitudes and therefore does not rely on a Breit--Wigner parametrization. It is also consistent with the chiral effective field theory prediction of Ref.~\cite{Epelbaum:2007sq}, which finds the $\Delta^0$ to be approximately $3\,\mathrm{MeV}$ heavier than the $\Delta^{++}$.

Our result for the width difference remains compatible with zero within uncertainties. After accounting for the different isospin normalization of the physical amplitudes, the doubly charged and neutral residue moduli come out compatible within uncertainties, while the residue phases show no statistically significant difference. The present model-independent determination confirms that $\pi^\pm p$ total cross-sections retain sensitivity to isospin-breaking effects in the $\Delta(1232)$ sector, while remaining compatible with existing determinations of the Breit-Wigner mass difference.

\section{Summary and conclusions}\label{sec:discussion}

We have presented a model-independent dispersive study of $\pi N$ forward dispersion relations and their analytic continuation to the complex plane, using almost exclusively $\pi^\pm p$ total cross-section data. 

To that end, we have revised the $\pi^\pm p$ total cross-section RPP dataset, implementing several improvements. These include the systematic use of the charged-pion mass when converting kinematic variables in $\pi^\pm p$ data, additional data points or sets that are absent from the RPP compilation, corrected central values and uncertainties, and purging sets to retain values from the latest reanalyses. For our final data fit with splines, we have not considered sets with disproportionately large uncertainties
or largely inconsistent with the rest.
We have also excluded experiments that lack details on their implementation of electromagnetic and isospin-breaking corrections, because we have implemented them in the final fit following~\cite{Tromborg:1976bi,Arndt:2003if,Arndt:2006bf}.

The imaginary parts of the forward amplitudes have been obtained from the total cross-section fit, by means of
the optical theorem up to $\sqrt{s}=3\,\mathrm{GeV}$. Above that energy, we have used Regge asymptotics. The corresponding real parts have been obtained through forward dispersion relations exploiting analyticity, unitarity, and crossing symmetry. The resulting forward amplitudes are fairly compatible with those obtained from the Roy--Steiner solution of~\cite{Hoferichter:2015hva} at low energies and from the SAID/GWU analysis~\cite{Workman:2012hx,SAID} up to $\sqrt{s}\simeq2.2\,\mathrm{GeV}$.

Analytically continuing the dispersive amplitudes into the complex-energy plane with the $\mathrm{FDR}_{C_N}$ method, we find four stable poles for each of the $I=3/2$ and $I=1/2$ isospin channels.  The lightest poles are unambiguously associated with the $\Delta(1232)$ and the Roper resonance $N(1440)$, respectively. However, although the other six singularities cannot be associated with individual resonances, we have shown that they are not artifacts of the approach. Instead, they are generated from the combined effect of multiple states.
We have first illustrated this mechanism using the SAID/GWU solution by reconstructing forward amplitudes from its partial-wave parametrization. The parameters of the poles appearing in the FDRs' analytic continuation indeed arise from the poles found in the underlying SAID/GWU partial-wave analysis. The FDR$_{C_N}$ method cannot always resolve the poles of individual resonances.

When using our model-independent fit to the revised total cross-section data and their uncertainties, the FDR$_{C_N}$ resolution is even worse. Still, the analytic structure of the forward amplitudes encodes resonance information beyond what is apparent in the data without relying on partial-wave analyses. First, we have been able to determine the parameters of the $\Delta(1232)$ and 
$N(1440)$ resonances. The appearance of a stable Roper pole is particularly noteworthy given that it is almost imperceptible in the total cross-section. We have thus provided the first determination of its parameters from $\pi N$ total cross-section data, independent of models and partial wave analyses.

Second, we have obtained sum rules relating the residues of the remaining six dispersive poles with those of the resonances that combine to generate them. From the total cross-section
data, we have thus extracted dispersive values for the sum of the residues of these constituent resonances.
We have checked that these sum rules are consistent with the established RPP values.

Finally, our dispersive analysis further yields a model-independent estimate of isospin-breaking effects in the $\Delta(1232)$ sector. Comparing the pole parameters extracted from the $D_+$ and $D_-$ amplitudes separately, we find
\begin{equation}
    M_{\Delta^0}-M_{\Delta^{++}}=(8.1\pm3.9)\,\mathrm{MeV},
\end{equation}
while the corresponding width and residue differences remain compatible with zero within uncertainties. Although less precise than Breit-Wigner determinations of the $\Delta^0-\Delta^{++}$ splitting, this result shows that $\pi^\pm p$ total cross-sections retain sensitivity to isospin-breaking effects when analyzed within a dispersive framework.

Taken together, these results demonstrate that $\pi^\pm p$ total cross-sections, once supplemented by analyticity, unitarity, crossing symmetry, and high-energy constraints, already encode a substantial amount of information on the baryon resonance spectrum. At the same time, the intrinsic limitations of forward amplitudes are manifest: they do not resolve the spin of the underlying states and cannot separate nearby resonances with identical isospin quantum numbers. Extending the framework to non-forward kinematics or incorporating polarization observables would allow individual partial waves to be resolved and provide access to the spin quantum numbers of the underlying resonances. The present work thus provides a model-independent dispersive determination of resonance pole parameters from total cross-section data and a benchmark for future dispersive studies of the baryon spectrum.

\begin{acknowledgments} 
We thank M. D\"oring, M. Hoferichter, B. Kubis, U.-G. Meißner, M.~M.~Pavan, I.~I.~Strakovsky, and R.~L.~Workman for discussions and clarifications.  This work is part of the Grant PID2022-136510NB-C31 funded by MCIN/AEI/ 10.13039/501100011033, it has also received funding from the European Union’s Horizon 2020 research and innovation program under grant agreement No.824093. P. R. is supported
by the MIU (Ministerio de Universidades, Spain) fellowship FPU21/03878, and J.R.E. by the Ram\'on y Cajal program (RYC2019-027605-I) of the Spanish MINECO. 
\end{acknowledgments}

\appendix

\bibliographystyle{apsrev4-2}
\bibliography{largebiblio.bib}
\end{document}